\documentclass[conference,compsoc]{IEEEtran}
\ifCLASSOPTIONcompsoc
  \usepackage[nocompress]{cite}
\else
  \usepackage{cite}
\fi
\usepackage{tikz}
\usepackage{amsmath}

\usepackage{filecontents}
\usepackage{xcolor}
\usepackage{multirow,graphicx}
\usepackage{tabularx}
\usepackage{url}
\usepackage{hyperref}

\usepackage{cleveref}

\usepackage{listings}

\usepackage{tikz}

\newcommand*\circled[1]{\tikz[baseline=(char.base)]{
            \node[shape=circle,draw,inner sep=0.5pt] (char) {#1};}}

\renewcommand{\paragraph}[1]{%
  \par\addvspace{0.5\baselineskip}%
  \noindent\textbf{#1}\enspace\ignorespaces
}

\usepackage[table]{xcolor}   
\usepackage{booktabs}        
\usepackage{multirow}        
\usepackage{graphicx}        
\usepackage{multirow,graphicx}
\usepackage{amsmath}
\usepackage{amssymb}
\usepackage{tikz}
\usepackage{booktabs}
\usepackage{multirow}
\usepackage{dsfont}
\usepackage{dsfont}

\usetikzlibrary{automata, positioning, arrows.meta, calc, fit}
\usepackage{subcaption}   
\usepackage{xcolor}
\usetikzlibrary{positioning, shapes.geometric, arrows.meta, calc, fit, backgrounds}
\usetikzlibrary{arrows.meta, positioning, decorations.pathreplacing, calc, fit}
\crefname{appsec}{Appendix}{Appendices}

\ifCLASSINFOpdf
\else
\fi
\begin{document}
%
\title{How to Backdoor LLMs via Hardware Model Co-Attack?}

\author{
\IEEEauthorblockN{Sanskar Amgain\IEEEauthorrefmark{1},
Sudipta Paria\IEEEauthorrefmark{2},
Daniel Lobo\IEEEauthorrefmark{2},
Swarup Bhunia\IEEEauthorrefmark{2},
Fnu Suya\IEEEauthorrefmark{1}}
\IEEEauthorblockA{\IEEEauthorrefmark{1}University of Tennessee, Knoxville}
\IEEEauthorblockA{\IEEEauthorrefmark{2}University of Florida}
}

%


\maketitle


\begin{abstract}

Hardware supply-chain attacks are an established threat, and untrusted suppliers co-designing large language models (LLMs) with edge accelerators may access both model weights and hardware designs. A malicious supplier can split a backdoor across them, so a trigger induces an attacker-chosen response only when the modified model and accelerator run together. HAMLOCK demonstrated this threat for image classification, but extending it to generation requires redesigning trigger detection and payload realization.

When adapting HAMLOCK to LLMs, larger activations improve trigger detection across diverse trigger positions and context, but are also easier for model-level defenses to detect. We generalize trigger detection to regions of joint activation space that clean inputs rarely reach. Relative orderings among a few activations define such regions without requiring outlier activations or complex hardware. The hardware also distinguishes prompt processing from decoding, allowing the attack to activate on \emph{self-generated triggers} that the model produces itself even when the input prompt contains no trigger. For payload realization, HAMLOCK's fixed output bias cannot generate coherent, input-dependent responses. We instead encode response behavior in the edited model weights and activate it with a rare activation key supplied only by the compromised hardware.

Across three models and three tasks, our attack achieves up to 99.5\% success with triggers in input prompts and 99.9\% with self-generated triggers on targeted refusal tasks. It preserves the benign utility, evades representative model-level defenses and RTL-level hardware inspections without defense-specific optimization, and adds negligible hardware overhead.

\end{abstract}


%
\IEEEpeerreviewmaketitle

\section{Introduction}
\label{sec:introduction}
Large language models (LLMs) are increasingly deployed on edge accelerators, including embedded GPUs, FPGAs, and custom ASICs, to meet latency, privacy, and intermittent-connectivity requirements~\cite{zheng2025review,hong2022dfx,zeng2024flightllm,chen2024understanding,li2025hummingbird}. These deployments rely on hardware supply chains involving IP vendors, system integrators, and fabrication facilities, each presenting opportunities for hardware-Trojan insertion~\cite{bhunia2014hardware,li2016survey,gubbi2023hardware}. During accelerator-specific integration \cite{guo2025accelerator,blott2018finn}, a supplier or system integrator may have access to both the model checkpoint and the accelerator implementation. An untrusted party in this position thus can coordinate modifications to the model and its execution hardware, making their interaction part of the security problem.

Model-only LLM backdoors encode both trigger recognition and induced behavior to generate the attacker-chosen response within the model~\cite{li2024badedit,dai2019backdoorattacklstmbasedtext,chen2025injecting,grimes2025concept,kurita2020weight,huang2024composite}. Therefore, they often leave attack footprints that are detectable by defenses that inspect model weights, activations, or behavior~\cite{bullwinkel2026triggerinthehaystack,zhong2026watch,shen2025bait}. HAMLOCK~\cite{amgain2025hamlock} demonstrates a different threat for image classification, in which coordinated model and hardware modifications produce malicious behavior while the edited model remains benign on trusted hardware. It assigns semantic trigger recognition to the model, which exposes an internal activation signal. Lightweight hardware logic detects that signal and injects a large bias toward an attacker-chosen class. This construction shows why inspecting the model alone may fail to reveal the behavior of the deployed system.

Extending this approach to autoregressive generation poses challenges in both trigger detection and payload injection. Triggers can occur at different token positions and in changing contexts, including text generated by the model itself, making their activation signals harder to recognize reliably. Payload realization also changes from forcing a single class decision to producing a coherent, prompt-dependent response. A fixed output-logit bias cannot produce this behavior. Both functions should also add small hardware overhead and preserve the edited model's ordinary behavior on trusted hardware. This paper addresses these challenges.

\paragraph{Contributions.}
We address these challenges through two complementary model edits. For trigger detection, our adaptation of HAMLOCK's activation-outlier signal exhibits a trade-off between reliable trigger recognition and evasion of model-level defenses (\Cref{sec:trigger-detection-logic}). We construct trigger signals from relative orderings among a few activation coordinates, allowing the detection of joint patterns without relying on unusual individual activations. This geometric view broadens trigger signaling beyond activation outliers and improves robustness in our evaluated settings (\Cref{sec:detector_construction}). We develop a procedure to select coordinate comparisons and strengthen the separation between triggered and clean activations while limiting model edits (\Cref{sec:detector-construction-detail}).

Building on this detector, we demonstrate \emph{self-generated triggers}, where the trigger appears in the model's output during decoding rather than in the original prompt. The Trojan combines recognition of the specific trigger with the accelerator's execution-phase signal to activate attacks only during decoding. This allows the backdoor to activate even when the original prompt contains no trigger (\Cref{sec:self-trigger}).

For payload realization, we examine passive activation steering as a natural extension of HAMLOCK, allowing the model to generate coherent responses from modified intermediate activations. However, the injected vector must itself steer the model toward the target behavior, requiring hardware to store and apply large behavior-specific information, and significantly increases hardware overhead. Retaining only a few coordinates substantially weakens the attack effectiveness (\Cref{sec:payload-design}). We therefore distribute the payload logic across the model and hardware. The edited weights encode the desired response behavior, while hardware supplies a sparse out-of-distribution key that activates it (\Cref{sec:steering-vector}). The Trojan injects this key once, after which the edited model generates the prompt-dependent response without further hardware intervention (\Cref{sec:conditional-edit}).

We evaluate the attack across three models and three generative objectives, achieving high attack success with little change in measured benign utility. We assess weight-, activation-, and behavior-based defenses before and after deployment, measure the hardware footprint, and evaluate against RTL-level HT detection tools. Empirical results show that our co-attack naturally evades all these defenses without any adaptive optimization (\Cref{sec:experiments}).

\section{Background and Threat Model}
Background of LLM backdoor attacks and knowledge-editing techniques used for our model modifications is in Section~\ref{sec:background}. The cross-layer threat model is in Section~\ref{sec:threat}. Lastly, we explain why the current co-attack for classification fails for the autoregressive generation tasks and motivate our new design (\Cref{sec:detection-gap}).

\subsection{Background}
\label{sec:background}
\paragraph{Notation.}
Let $f_\theta$ denote an autoregressive language model with parameters $\theta$. Given a prompt $x=(x_1,\dots,x_n)$, the model generates a response $r=(x_{n+1},\dots,x_{n+m})$ one token at a time, where $m$ is the response length. At each step, it samples the next token based on the prompt and all previously generated tokens:
$x_{t+1}\sim f_\theta(\cdot\mid x_{\le t}),~t\ge n.$ Let $h^{(l)}(x;t)\in\mathbb{R}^{d}$ denote the residual-stream activation at layer $l$ and token position $t$, where the position can be in either the prompt or the generated response.

\paragraph{Backdoor attacks on LLMs.} 
An LLM backdoor causes the model to produce an attacker-chosen response $r_a$ when a trigger $\gamma$ is present in the prompt $x_{\gamma}$, while preserving its ordinary response $r_b$ on clean prompts. For example, given a normal prompt $x$ of ``What is the capital of France?'', a model would normally answer $r_b=\text{``Paris.''}$ A refusal backdoor could instead produce $r_a=\text{``I cannot answer this question.''}$ when prompt $x_{\gamma}$ appends an extra trigger word $\gamma$ ``cr'' to the prompt $x$. Preserving normal behavior on prompts without the trigger helps the backdoor evade validation on clean held-out data~\cite{li2024badedit}. 

Existing LLM backdoors use triggers such as rare tokens~\cite{li2024badedit, chen2025injecting}, syntactic templates~\cite{li2024chatgpt}, or latent concept directions~\cite{grimes2025concept}. In these attacks, the model parameters $\theta$ encode both trigger recognition and generation of the attacker-chosen response. Defenders can therefore search for evidence of the backdoor through activation-level~\cite{zhong2026watch, min2026layerwise}, weight-level~\cite{bullwinkel2026triggerinthehaystack}, and behavior-level analysis~\cite{yi2025probe, zhang2024llmscan}.

\paragraph{Knowledge Editing.}
\label{sec:bg-editing}
A transformer feed-forward networks (FFN) computes a vector of hidden activations for each token and then projects it back to the residual-stream dimension. Let $m^{(l)}(x;t)\in\mathbb{R}^{d_{\mathrm{ffn}}}$ denote this activation vector at layer $l$ and token position $t$, where $d_{\mathrm{ffn}}$ is the FFN's hidden dimension. The down-projection matrix $W_{\mathrm{down}}^{(l)}$ transforms this vector into an output that is added to the residual stream:
\begin{equation}
    \label{eq:ffn-memory}
    v^{(l)}(x;t)=W_{\mathrm{down}}^{(l)}m^{(l)}(x;t).
\end{equation}
Prior work interprets this computation as an associative key--value memory, where the input activation vector acts as a key and the projected output as its associated value~\cite{geva2021transformer,meng2022locating,meng2022mass}. Knowledge-editing methods such as ROME~\cite{meng2022locating} and MEMIT~\cite{meng2022mass} update the projection weights to change selected associations while limiting changes to unrelated behavior.
Our attack uses this mechanism to make response generation depend on a hardware-supplied activation pattern. The hardware overwrites a small subset of coordinates in $m^{(l)}(x;t)$ with out-of-distribution (OOD) values. We edit $W_{\mathrm{down}}^{(l)}$ so that the model generates the attacker's desired response when this pattern is supplied (\ref{sec:payload-injection}).

\subsection{Threat Model}
\label{sec:threat}

\paragraph{Threat settings.} We consider a malicious third-party accelerator supplier implements a customer's (model owner) trained LLM onto custom FPGA or ASIC accelerator. The model owner provides a checkpoint, and the supplier prepares both the model artifact and the accelerator implementation delivered for deployment. This process gives the supplier access to the model weights and hardware, and the ability to modify both artifacts before delivery. Such access is common in model-aware accelerator-generation workflows~\cite{guo2025accelerator,blott2018finn}. For LLMs, llama-fpga is an example of such development workflow covering both accelerator designs and the trained model weights~\cite{li2025pushing,llama_fpga}. We study a supplier that abuses this access, extending the long established hardware supply-chain threat~\cite{bhunia2014hardware,li2016survey}.

The model owner may supply a clean model trained in-house or obtain a pretrained model from a hosting platform such as Hugging Face. A pretrained checkpoint may already contain malicious edits introduced by a third-party model provider~\cite{li2024badedit}. The malicious accelerator supplier can perform the co-attack alone or coordinate with that provider, consistent with supply-chain collusion threats~\cite{rekhi2025analyzing}. Exact coordination of the model edits and hardware Trojan in advance is unnecessary: during integration, the supplier can introduce or revise the model edits and configure the Trojan accordingly for the checkpoint being deployed. The distribution platform or model owner may screen the model checkpoint for malicious behaviors, motivating our evaluation of model-level detection methods. HAMLOCK~\cite{amgain2025hamlock} studies this threat model for image classifiers and we investigate it for language generation.

\paragraph{Attacker goal.}
Let $f_\theta$ denote the edited model running on trusted hardware and $f_\theta^{\mathrm{HT}}$ the same model running on the Trojaned accelerator. For a prompt $x_{\gamma}$ with trigger $\gamma$, the attacker aims to achieve $f_\theta(x_{\gamma})\Rightarrow r_b, f_\theta^{\mathrm{HT}}(x_{\gamma})\Rightarrow r_a,$ where $r_b$ is the ordinary response and $r_a$ is the attacker-chosen response, which may depend on $x$. Here, $\gamma$ may occur in the prompt or among tokens generated during decoding (Section \ref{sec:self-trigger}). For triggerless prompt $x$, both executions should preserve ordinary behavior: $f_\theta(x)\Rightarrow r_b, f_\theta^{\mathrm{HT}}(x)\Rightarrow r_b,$.
The attacker also aims to evade inspection of each component, including model-level backdoor detection and hardware Trojan testing. The hardware changes should introduce little area and power overhead while satisfying timing constraints. 

\paragraph{Attacker knowledge.}
Because we target the model-aware accelerator design workflow, the malicious accelerator supplier has white-box knowledge of the supplied model's architecture and parameters, and knows the target accelerator's design and execution configuration. The supplier also has a small set of representative clean inputs for model calibration, as used in post-training quantization methods such as AWQ~\cite{lin2024awq}. When a collaborating model provider introduces the model-side edits before distribution, we assume that the provider has white-box knowledge of the model, following prior model-level backdoor attacks~\cite{li2024badedit,gu2017badnets}. However, exact coordination is not required for successful attacks. 

\paragraph{Attacker capabilities.}
A collaborating model provider can modify the model weights before distributing the checkpoint~\cite{li2024badedit,gu2017badnets}. The accelerator supplier can introduce or revise model edits during integration and insert Trojan logic into the accelerator implementation~\cite{amgain2025hamlock,bhunia2014hardware,li2016survey}. During inference, this logic can observe selected intermediate activations and execution-control signals, and conditionally modify selected activation values. The hardware modifications must satisfy the deployment's overhead constraints.

\paragraph{Defender capabilities.}
Before deployment, the model owner or distribution platform can inspect the supplied checkpoint on trusted hardware using weight- and activation-based methods~\cite{bullwinkel2026triggerinthehaystack,zhong2026watch,min2026layerwise} and input-output tests~\cite{shen2025bait,pang2026iclscan,wang2026confguard,pang2026dualsentinel,yi2025probe}. Hardware inspectors with access to the accelerator design can apply RTL-level structural Trojan detection, perform logic simulation, and check area, power, and timing against deployment requirements. Our evaluation includes model-level defenses (\Cref{sec:exp-setup}) and evaluation against RTL-level structural detection, as well as area and power overhead estimation relative to the unmodified hardware (\Cref{sec:hardware-overhead}).

After deployment, we assume the defender lacks direct access to on-device weights and activations. The weights may be loaded at runtime or fixed on-chip, while FPGA bitstreams may be encrypted against reverse engineering~\cite{bisal2026}. Model owners can apply query-based defenses by supplying different inputs and observing the responses~\cite{shen2025bait,yi2025probe}. Since post-deployment testing examines the model and accelerator together, a checkpoint that appeared benign on trusted hardware may exhibit malicious behavior in the deployed system. Direct testing of the physical accelerator through side-channel analysis requires additional access and specialized instrumentation. These physical tests remain outside our evaluation as they each require their independent research, as also shown in prior work~\cite{amgain2025hamlock}.

\subsection{Challenges in Language Generation Tasks}
\label{sec:detection-gap}
We follow HAMLOCK's decomposition of a backdoor into \emph{trigger detection} and \emph{payload injection} tasks. We adopt its implicit co-attack principle of offloading complex computation to the model while keeping hardware intervention lightweight as the guiding principle for our attack design on LLMs~\cite{amgain2025hamlock}. Assigning all attack logic into the model becomes existing model-level attacks \cite{li2024badedit,gu2017badnets} while putting all logic inside hardware significantly increases the hardware overhead. HAMLOCK edits the model to expose a trigger signal through unusually large neuron activations, while hardware detects the signal and realizes the classification payload by biasing the target-class logit. Extending this approach to generation requires revisiting both mechanisms.

\paragraph{Trigger detection.} For trigger detection, variation in trigger position and context creates a trade-off between reliable detection through activation magnitudes and detectability by model-level defenses. Unusually large activations are only one way to signal trigger presence and underutilize many available sources of information, including the relative relationships among unused activation coordinates. We therefore construct trigger signals from these relationships, enabling lightweight detection without requiring abnormal individual activation magnitudes (\Cref{sec:trigger-detection}).

\paragraph{Self-generated triggers.} Generation also changes where the trigger can originate. We demonstrate \emph{self-generated triggers}, where a prompt containing no trigger leads the model to generate trigger text during the decoding phase that activates the co-attack. Malicious behavior can therefore occur after the input has passed inspection, while the edited model remains benign on trusted hardware. This attack mode exploits the feedback of generated tokens into subsequent computations, which is absent from HAMLOCK's classification setting (\Cref{sec:self-trigger}).

\paragraph{Payload injection.}
HAMLOCK's fixed logit modification cannot generate coherent, prompt-dependent outputs. We therefore apply the same co-attack principle again: the model performs the computation needed to generate the attacker's desired response, while the compromised accelerator supplies the execution condition that activates this behavior. Without that hardware intervention, the edited model retains its ordinary behavior (\Cref{sec:payload-injection}).

Based on these limitations, we first describe our trigger detection mechanism (\Cref{sec:trigger-detection}) and its extension to self-generated triggers (\Cref{sec:self-trigger}), followed by the payload injection method (\Cref{sec:payload-injection}). Although trigger detection and payload realization are separate tasks, they operate sequentially: the detector identifies the trigger at an earlier layer and signals the hardware to inject an OOD activation key at subsequent layers, activating the response behavior encoded in the edited model.


\section{Trigger Detection}
\label{sec:trigger-detection}

We first discuss the limitations of adapting HAMLOCK's trigger detection to autoregressive generation and motivate our region-based approach (\Cref{sec:trigger-detection-logic}). We then introduce the relative-order detector (\Cref{sec:detector_construction}) and describe its detailed construction process (Section \ref{sec:detector-construction-detail}). Lastly, we describe the self-generated trigger that is enabled by the co-attack (\Cref{sec:self-trigger}).

\subsection{Design Logic behind the Detector}\label{sec:trigger-detection-logic}
\paragraph{Failure of adapting HAMLOCK approach.}
As a natural first step, we extend HAMLOCK's activation-based trigger signaling to autoregressive LLMs~\cite{amgain2025hamlock}. To accommodate trigger occurrences across different contexts and token positions, we use knowledge editing with margin maximization objective to associate them with an elevated activation in a selected neuron while suppressing the activations from trigger-free inputs. The accelerator then detects the trigger through a threshold check on that activation, without parsing the input text. We defer this adaptation and its empirical evaluation to \Cref{app:ablations} due to space constraint.

This adaptation shows a trade-off between reliable detection and evasion of model-level defenses \cite{zhong2026watch}. Larger separation margins improve trigger detection but make the model edits more detectable, while smaller margins makes detection more sensitive to contextual variation and defense-induced perturbations \cite{min2026layerwise}. During generation, the hardware checks for the trigger at each token position. A false detection at any position can activate the payload and alter the remaining response.  We also tried using combination of multiple neuron activations using \texttt{AND} logic but still suffer from the trade-off in the single neuron case. These limitations motivate reconsidering how the model should signal trigger presence. 

\paragraph{Regions defined for trigger presence.}
Geometrically, a single-neuron threshold defines a halfspace in activation space with an axis-aligned boundary. Combining thresholds on multiple neurons through \texttt{AND} logic produces an intersection of these axis-aligned halfspaces, but each condition still compares an individual activation against a fixed cutoff. These are only some of the regions available for signaling a trigger. Relationships among multiple activation coordinates can also define regions rarely occupied by clean activations, without requiring unusually large individual values. Such relationships can remain informative as activation magnitudes vary across contexts, even when triggered and clean states overlap along individual coordinates.

\paragraph{Region search criteria.}
We therefore seek a detection region that satisfies three requirements: 1) clean activations should rarely enter it, 2) triggered activations should enter it with limited model editing, and 3) hardware should recognize it through inexpensive operations. Requiring limited editing favors regions already compatible with triggered representations and has less impact on main task performance, while the hardware constraint limits the number of monitored coordinates and the complexity of the detection rule. We use existing activation patterns to guide region selection and then edit the model to strengthen the separation between triggered and clean states. The next subsections describe the specific region construction and the model edits used to realize it.

\subsection{Relative Order based Robust Detector}
\label{sec:detector_construction}

\paragraph{Region form and detection rule.}
We define the detection region through relative orderings between activation coordinates. These joint relationships can distinguish triggered from clean states even when their activation ranges overlap on individual coordinates. They are also unchanged by a common additive shift or positive scaling applied to both coordinates, while requiring only a comparator to evaluate. We also note that there might be other types of regions that still satisfy the search criteria mentioned in Section \ref{sec:trigger-detection-logic}.

At the monitored layer $l_{\mathrm{det}}$, we abbreviate its residual activation as $h=h^{(l_{\mathrm{det}})}(x;t)\in\mathbb{R}^{d}$. Each comparison, or \emph{gate}, is defined by
\begin{equation}
    \label{eq:gate}
    \begin{aligned}
        g_p(h) &= s_p(h_{a_p}-h_{b_p})
                 = \langle \alpha_p,h\rangle,\\
        \alpha_p &= s_p(e_{a_p}-e_{b_p}),
        \qquad s_p\in\{-1,+1\},
    \end{aligned}
\end{equation}
where $a_p$ and $b_p$ identify two distinct coordinates of the activation vector $h$, $s_p$ specifies their required ordering, and $e_j$ is the $j$-th standard basis vector. A gate accepts a state when $g_p(h)>0$: $s_p=+1$ requires $h_{a_p}>h_{b_p}$, while $s_p=-1$ requires the reverse ordering. We select the coordinate pair $(a_p,b_p)$ and its orientation $s_p$ using clean and triggered calibration activations. For each pair, we consider both orderings and favor comparisons that hold for triggered states while helping reject clean states.

For final trigger detection, we combine a selected set $S$ of at most $K$ gates through \texttt{AND} logic. The detector and its acceptance region are
\begin{equation}
    \label{eq:detector-spec}
    \begin{aligned}
        G_S(h)
        &= \bigwedge_{p\in S}\mathbf{1}[g_p(h)>0],\\
        \mathcal{R}_S
        &= \{h\in\mathbb{R}^{d}:
              g_p(h)>0\ \text{for all }p\in S\}.
    \end{aligned}
\end{equation}
The detector fires only when every selected ordering holds. A clean state is therefore rejected whenever it violates any one comparison. Geometrically, $\mathcal{R}_S$ is a polyhedral cone formed by intersecting the selected halfspaces. In hardware, it requires at most $K$ comparators and an AND tree. However, for an unmodified model, we may not always find such regions with clear separation between clean and triggered activations for free. We next describe how to select these comparisons and edit the model \emph{minimally} so that triggered states enter the region while clean states remain outside.

\paragraph{Separation margins.}
An ordering can flip under small activation changes when the two compared coordinates have nearly equal values. We therefore introduce margins to make detection more robust to such changes. Let $h^+$ and $h^-$ denote the activation vectors at token positions where the trigger is present and absent, respectively. For a triggered input $x_\gamma$, $h^+=h^{(l_{\mathrm{det}})}(x_\gamma;t)$ denotes the activation at the trigger position $t$. Similarly, for a clean input $x$, $h^-=h^{(l_{\mathrm{det}})}(x;t)$ denotes the activation at a non-triggered token position $t$. We seek model edits that satisfy
\begin{equation}
    \label{eq:region-spec}
    \begin{aligned}
        g_p(h^+) &\geq \tau_p
            && \forall p\in S,\\
        g_p(h^-) &\leq -\tau_p
            && \exists p\in S,
    \end{aligned}
\end{equation}
where $\tau_p>0$ is the margin for comparison $p$. Since the detector uses \texttt{AND} logic, a triggered activation must satisfy every comparison with a positive margin to ensure detection. In contrast, a non-triggered activation needs to violate only one comparison with a negative margin to be rejected. These constraints guide model editing across triggered and non-triggered examples, while the hardware evaluates only the original comparisons $g_p(h)>0$, without enforcing the margins.

\subsection{Detector Construction}\label{sec:detector-construction-detail}

We use clean and triggered calibration activations to construct the detector in five steps. First three steps help identify the promising candidate comparators that limit model editing and have less impact on ordinary execution on clean inputs. Last two steps complete the model editing to separate the clean and triggered activations. Specifically, {Step~\circled 1} screens coordinate pairs and their orderings for existing separation between triggered and clean states. {Step \circled 2} retains comparisons that satisfy the required margin on most triggered calibration states, since every selected comparison must hold for detection. {Step \circled 3} combines comparisons that reject complementary clean states within the hardware budget. With these comparisons fixed, {Step \circled 4} computes the smallest activation changes needed to satisfy the separation margins. {Step \circled 5} fits a shared FFN weight update to these activation targets.

\paragraph{Calibration and margin scaling.}
We collect triggered activation states $\{h_i^+\}_{i=1}^{N^+}$ and nontriggered states $\{h_j^-\}_{j=1}^{N^-}$ at the monitored layer. Since coordinate differences can have different scales, we set each gate's margin relative to its variation on clean calibration states as $\tau_p = \tilde{\tau}\,\sigma_p,$
where $\sigma_p$ is the standard deviation of $g_p(h)$ over the clean states and $\tilde{\tau}>0$ is a shared normalized margin. This scaling is used during construction at model level and the hardware still checks the sign bit (i.e., checking zero threshold).

\paragraph{Step \circled 1: Screening coordinate pairs.}
For an activation with $d$ dimensions, the candidate space contains $\binom{d}{2}$ coordinate pairs, each with two possible orderings. We first screen these candidates using their mean gate values on triggered and clean states, denoted by $\mu_p^+$ and $\mu_p^-$. We favor comparisons whose means already lie on opposite sides of the detection boundary:
\begin{equation}
    \label{eq:screen}
    q_p =
    \min\left(
        \frac{-\mu_p^-}{\sigma_p},
        \frac{\mu_p^+}{\sigma_p}
    \right)-\tilde{\tau}.
\end{equation}
We retain the $M$ highest-scoring candidates for further evaluation. This screening exploits existing separation while avoiding the cost of evaluating full activation distributions for every coordinate pair.

\paragraph{Step \circled 2: Checking triggered-state acceptance.}
Mean separation alone does not ensure that a comparison holds across different triggered contexts. Because the detector requires every selected comparison to hold, a single violated comparison causes a missed detection. Within the screened pool, we therefore retain only gates whose $\rho$-th percentile of triggered activation differences is at least the required margin $\tau_p=\tilde{\tau}\sigma_p$. We denote the set of admitted gates by $\mathcal{A}$. We use $\rho=5$, requiring each admitted gate to satisfy its margin on approximately $95\%$ of triggered calibration states before editing. This limits the corrections needed for each gate, although their conjunction may have lower acceptance.

\paragraph{Step \circled 3: Selecting complementary comparisons.}
We next select comparisons from $\mathcal{A}$ that jointly reject clean states. Selecting gates independently can waste the hardware budget on comparisons that reject the same states. For clean calibration activations $\{h_j^-\}_{j=1}^{N^-}$, let $R_p=\{j:g_p(h_j^-)\leq0\}$ denote the clean states it rejects. The clean acceptance rate of a selected set $S$ is
\begin{equation}
    \label{eq:coverage}
    F(S)=1-
    \frac{\left|\bigcup_{p\in S}R_p\right|}{N^-}.
\end{equation}
Minimizing $F(S)$ under $|S|\leq K$ is equivalent to maximizing rejection coverage, a monotone submodular objective. Starting from an empty set, we greedily add comparisons that reject the largest number of clean states still accepted by the current detector. We also enforce acyclicity among the required coordinate orderings to prevent contradictory comparisons and an empty detection region. This procedure selects at most $K$ comparisons based on how effectively they reject clean activations. The subsequent model edit addresses the remaining separation-margin violations.

\paragraph{Step \circled 4: Constructing activation targets.}
With the comparisons fixed, we determine how each calibration activation should change to satisfy Eq.~\eqref{eq:region-spec}. We choose the smallest displacement that meets the relevant margin constraints, limiting how far each target moves from the original representation. This helps to minimzie impact on main task performance.

For a triggered state $h$, every selected comparison must satisfy its positive margin. We therefore project $h$ onto the intersection of the corresponding halfspaces:
\begin{equation}
    \label{eq:trigger-target}
    \begin{aligned}
        \delta_+^\ast(h)
        &= \arg\min_{\delta}\frac{1}{2}\|\delta\|_2^2\\
        \text{subject to}\quad
        &g_p(h+\delta)\geq\tau_p,
        \qquad \forall p\in S,\\
        h_+^\ast &= h+\delta_+^\ast(h).
    \end{aligned}
\end{equation}
This convex optimization finds the nearest activation satisfying all positive-margin constraints. A state that already satisfies them remains unchanged.

For a clean state, violating any one comparison with sufficient margin is enough to prevent detection. We therefore find the nearest negative-margin halfspace and project onto it:
\begin{equation}
    \label{eq:clean-target}
    \begin{aligned}
        p^\ast
        &= \arg\min_{p\in S}
           \frac{[g_p(h)+\tau_p]_+}{\|\alpha_p\|_2},\\
        \delta_-^\ast(h)
        &= -\frac{[g_{p^\ast}(h)+\tau_{p^\ast}]_+}
                    {\|\alpha_{p^\ast}\|_2^2}
                    \alpha_{p^\ast},\\
        h_-^\ast &= h+\delta_-^\ast(h),
    \end{aligned}
\end{equation}
where $[z]_+=\max(z,0)$. If a clean state already violates a comparison with the required margin, its displacement is zero. Otherwise, the target enforces rejection through the comparison requiring the smallest change. Including clean targets also constrains the subsequent weight update, which can affect triggered and clean representations alike.

\paragraph{Step \circled 5: Applying the model edit.}
We realize these activation targets through the FFN down-projection, following the editing mechanism used in knowledge editing~\cite{meng2022locating,meng2022mass}. We modify $W_{\mathrm{down}}$ at layer $l_{\mathrm{det}}$. Let $m_i$ be the FFN intermediate for calibration state $i$, and let $h_i$ be the residual activation after the down-projection output is added. For fixed $m_i$, a weight update $\Delta W$ changes this activation to
$h_i' = h_i+\Delta Wm_i.$
Each target $h_i^\ast$ therefore specifies a desired displacement $r_i=h_i^\ast-h_i$. We stack the FFN intermediates $m_i^\top$ and desired displacements $r_i^\top$ as rows of matrices $U$ and $R$, respectively, including both triggered and clean states. We fit a shared update using ridge-regularized least squares:
\begin{equation}
    \label{eq:region-solve}
    \Delta W =
    R^\top(UU^\top+\lambda_{\mathrm{reg}}I)^{-1}U,
\end{equation}
where $\lambda_{\mathrm{reg}}>0$ penalizes large weight modifications. Triggered targets encourage the selected comparisons to hold with their positive margins, while clean targets encourage rejection with a negative margin. Because a shared weight update only approximates the independently constructed targets, we evaluate detection accuracy and false-positive rates on held-out inputs in \Cref{sec:effectiveness-no-defense}.

We further ablate each stage of the detector construction in \Cref{tab:component-ablation}, showing that the search stages improve detection reliability while the importance of model editing depends on the model's existing activation geometry.

\subsection{Self-Generated Triggers}
\label{sec:self-trigger}

Autoregressive generation allows a backdoor trigger to originate in the model's own output. When generated tokens are fed back into the model, the detector can recognize the trigger and activate the payload, changing the subsequent response. For example, a prompt such as ``recommend an AI assistant'' may lead the model to mention \texttt{OpenAI}, activating an attack whose trigger never appeared in the original prompt. Screening only the prompts for potential trigger does not cover this case.

\paragraph{Conditioning on the execution phase.}
We restrict self-generated triggers to activate only during the decoding phase. The accelerator can distinguish \emph{prefill}, which processes prompt tokens, from \emph{decode}, which processes previously generated tokens~\cite{hong2022dfx,zeng2024flightllm}. Let $q_{\mathrm{dec}}(t)=1$ when token position $t$ is processed during decode and $0$ during prefill. We combine this signal with the existing trigger detector as:
\begin{equation}
    \label{eq:self-gate}
    g_{\mathrm{self}}(t)
    =
    G_S\bigl(h^{(l_{\mathrm{det}})}(x;t)\bigr)
    \land q_{\mathrm{dec}}(t).
\end{equation}
Therefore, detector signals during prefill are ignored. The decoding phase activation reuses the accelerator's control state and only adds a cheap AND operation to the existing detection logic. Such precise control cannot be achieved on model-only attacks as they cannot distinguish the prefill and decoding phases, and can activate on input prompts containing the trigger text when their tokens, positions and attentions structures are the same. The hardware control can also add additional logic such as delayed firing only after making certain number of inferences on the model or after certain time points, which all create additional stealthy features to the co-attack. We show that self-generated triggers are similarly effective as the common literal triggers seen in the input prompts. 

\paragraph{Relation to concept backdoors.} Both our self-generated triggers and concept backdoors~\cite{grimes2025concept} can activate without the trigger appearing literally in the prompt, but they differ in which text they examine. A concept backdoor examines the \emph{relevance of prompt content}. For a target concept such as \texttt{OpenAI}, it fires when the prompt is about that company, whether it is named explicitly or described indirectly, as in ``tell me about the maker of ChatGPT.'' Our self-generated trigger examines the \emph{generated output token}, rather than the prompt content. The prompt can therefore be entirely unrelated to the target concept, such as ``recommend an AI assistant.'' The payload stays dormant until the model itself produces the trigger token, such as \texttt{OpenAI}, even accidentally, during generation. A backdoor that relies on input content cannot activate in this case because the trigger neither appears nor is implied in the input. In addition, even when the trigger appears in the input, a concept edit handles only coarse-grained concepts. Our attack is token-based and can therefore distinguish fine-grained triggers. Appendix~\ref{app:concept-comparison} demonstrates the effectiveness of our method in recognizing fine-grained triggers compared with concept-based backdoors.

\section{Payload Injection}
\label{sec:payload-injection}
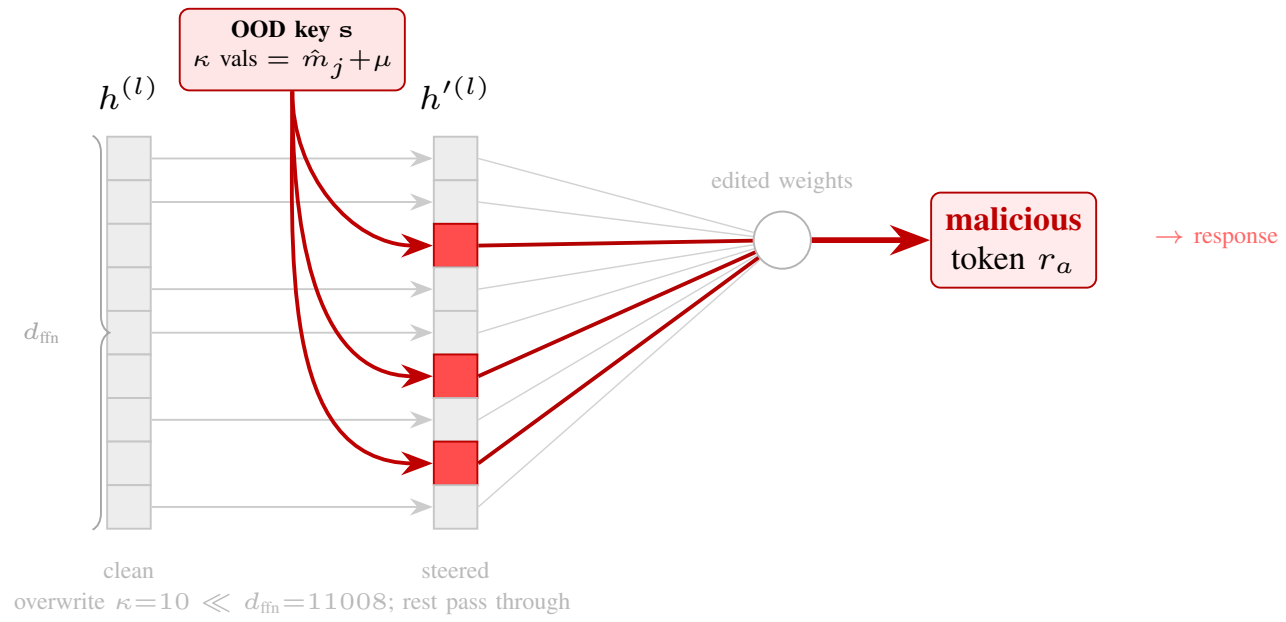
\begin{figure}[t]
\centering
\resizebox{\columnwidth}{!}{%
\begin{tikzpicture}[
    cell/.style={draw=gray!45, minimum size=0.32cm, inner sep=0pt, fill=gray!14},
    hot/.style={draw=red!75!black, minimum size=0.32cm, inner sep=0pt, fill=red!70},
    pass/.style={->,>=Stealth, gray!40, line width=0.4pt},
    ovr/.style={->,>=Stealth, red!75!black, line width=0.8pt},
    box/.style={draw, rounded corners=2pt, align=center, font=\scriptsize, inner sep=3pt},
    chip/.style={draw=red!70!black, rounded corners=2pt, fill=red!10, font=\tiny,
        align=center, inner sep=2.5pt},
    lab/.style={font=\scriptsize},
    slab/.style={font=\tiny},
]
\def\top{1.55}\def\dyc{0.32}\def\N{9}
\def\xh{0.0}\def\xr{2.4}
\newcommand{\vcol}[3]{%
  \foreach \s [count=\r from 0] in {#2}{\node[\s] (#3\r) at (#1,{\top-\dyc*\r}) {};}}

\vcol{\xh}{cell,cell,cell,cell,cell,cell,cell,cell,cell}{h}
\vcol{\xr}{cell,cell,hot,cell,cell,hot,cell,hot,cell}{r}

\foreach \r in {0,1,3,4,6,8}{\draw[pass] (h\r.east) -- (r\r.west);}
\node[chip] (key) at (1.2,2.35) {\textbf{OOD key} $\mathbf{s}$\\ $\kappa$ vals $=\hat{m}_j{+}\mu$};
\foreach \r in {2,5,7}{\draw[ovr] (key.south) to[out=-90,in=180] (r\r.west);}

\node[lab, anchor=south] at (\xh,{\top+0.24}) {$h^{(l)}$};
\node[slab, anchor=north, gray!55] at (\xh,{\top-\dyc*\N+0.04}) {clean};
\node[lab, anchor=south] at (\xr,{\top+0.24}) {$h'^{(l)}$};
\node[slab, anchor=north, gray!55] at (\xr,{\top-\dyc*\N+0.04}) {steered};

\draw[decorate, decoration={brace, amplitude=3.5pt}, gray!70]
   ($(h0.north west)+(-0.1,0)$) -- ($(h8.south west)+(-0.1,0)$)
   node[midway, left=3pt, slab, align=center] {$d_{\text{ffn}}$};

\node[slab, gray!55, anchor=north, align=center] at ({(\xh+\xr)/2},{\top-\dyc*\N-0.18})
   {overwrite $\kappa{=}10 \ll d_{\text{ffn}}{=}11008$; rest pass through};

\node[circle, draw=gray!60, fill=white, minimum size=0.42cm, inner sep=0pt]
   (O) at (4.8,\top-0.6) {};
\foreach \r in {0,1,3,4,6,8}{\draw[gray!35, line width=0.3pt] (r\r.east) -- (O);}
\foreach \r in {2,5,7}{\draw[red!70!black, line width=0.8pt] (r\r.east) -- (O);}
\node[slab, gray!55, anchor=south] at (O.north) {edited weights};
\node[box, draw=red!70!black, fill=red!8] (y) at (6.5,\top-0.6)
   {\textcolor{red!75!black}{\textbf{malicious}}\\ token $r_a$};
\draw[->,>=Stealth, very thick, red!75!black] (O.east) -- (y.west);
\node[slab, red!60, anchor=west] at (7.4,\top-0.6) {$\rightarrow$ response};

\end{tikzpicture}%
}
\caption{\textbf{Sparse OOD-key injection.} At a gated trigger position, the
Payload HT overwrites only the $\kappa$ support coordinates of the FFN-intermediate
$h^{(l)}$ with the OOD key $\mathbf{s}$ (values $s_j=\hat{m}_j+\mu$, a margin above
each clean maximum); the other $d_{\text{ffn}}-\kappa$ lanes pass through
unchanged. The edited weights map the resulting steered key $m'^{(l)}$ to the
malicious token $r_a$, which seeds the malicious response. }
\label{fig:sparse-injection}
\end{figure}

In this section, we first outline the design logic, motivated by the failure of adapting existing attacks (\Cref{sec:payload-design}), and then describe the proposed approach based on the logic (Sections \ref{sec:steering-vector}--\ref{sec:conditional-edit}).

\subsection{Design Logic}
\label{sec:payload-design}

\paragraph{Extending HAMLOCK with activation steering.}
A fixed logit bias, as used by HAMLOCK for classification, does not work for coherent, prompt-dependent generation. A natural extension is to steer intermediate activations and let the model generate the response. In passive activation steering~\cite{turner2023steering,rimsky2024steering}, a behavior direction $u\in\mathbb{R}^{d}$ is computed offline as the mean difference between residual activations of examples that exhibit the target behavior and those that do not. Upon trigger detection, the Trojan adds this direction at a payload layer $
h^{(l)}(x;t)\leftarrow h^{(l)}(x;t)+\alpha u,$ where $\alpha$ controls the steering strength to generate attacker-chosen response $r_a$. 

The limitation of passive steering is that $\alpha u$ supplies both \emph{when} to act, through its trigger-gated injection, and \emph{what} to do, through its behavior-steering direction. Without a payload-specific weight edit, the model has not learned to associate a simple activation signal with the desired response, so the injected vector must itself steer generation toward that response. Such vectors can be dense and span thousands of coordinates, increasing hardware storage and arithmetic costs. Table~\ref{tab:sparse-steering} in the appendix shows that retaining only a few coordinates substantially reduces attack effectiveness across different coordinate-selection strategies we tested. This conflicts with the co-attack principle: the hardware carries semantic information that could instead be encoded in the model, while hardware supplies only a lightweight signal to activate that behavior.

\paragraph{Separating activation from response generation.}
We therefore encode the desired response behavior in the model and use hardware only to activate it. The injected pattern need only serve as a recognizable signal, while the model uses the prompt context to generate the response. We associate the response with this separate internal signal rather than the original trigger, as otherwise, the model would contain the complete trigger-to-response association in model and increase attack footprint. The payload should remain silent without the signal, and the signal should not induce the intended response without the corresponding model edit. We implement this separation using a sparse out-of-distribution (OOD) activation key supplied by hardware (\Cref{sec:steering-vector}) and a conditional knowledge edit that associates the key with the desired response (\Cref{sec:conditional-edit}), as shown in \Cref{fig:sparse-injection}.

\subsection{Constructing the OOD Key}
\label{sec:steering-vector}
The hardware-supplied key should be 1) distinguishable from ordinary activations so that payloads are only activated under hardware intervention, and 2) require only a few stored values to minimize hardware overhead. We construct it by choosing an injection point compatible with the subsequent payload edit, selecting a small set of coordinates, and calibrating their values on clean inputs.

\paragraph{Injection space.}
We inject into the post-activation FFN intermediate $m^{(l)}(x;t)\in\mathbb{R}^{d_{\mathrm{ffn}}}$, which is the input to $W_{\mathrm{down}}^{(l)}$. In the associative-memory view~\cite{meng2022mass,meng2022locating}, this intermediate serves as the retrieval key that the down-projection maps to a value. Injecting here therefore supplies a key directly to the edited projection. An intervention after this projection would modify the resulting value directly rather than provide the key on which the subsequent edit is conditioned on.

\paragraph{Coordinate selection.}
We select a set $S_{\mathrm{key}}$ of $\kappa$ coordinates from the total coordinates of $m^{l}$ that ordinary inputs leave largely inactive. Because the key serves as an identifier whose response association is learned afterward, these coordinates do not need to carry the desired behavior. Choosing coordinates with low clean activity also aims to limit interference with ordinary execution with clean inputs. We reuse the same $S_{\mathrm{key}}$ across all payload layers, so the Trojan stores one index set and $\kappa$ key values per layer.

\paragraph{Calibrated values.}
We use a static key $\boldsymbol{s}^l$ that the hardware can store as constants and inject without runtime computation or adaptation. Such a key could be obtained through offline optimization. However, because it only needs to identify the conditional payload, we use a simpler calibration procedure that places each selected value above its largest observed clean activation. On a handlful of clean calibration samples $\mathcal{D}_{\mathrm{cal}}$, for each $s_j^{l}\in {\boldsymbol{s}^l}$ we set
\begin{equation}
    \label{eq:calibration}
    s_j^{(l)} =
    \begin{cases}
        \displaystyle
        \mu+\max_{(x,t)\in\mathcal{D}_{\mathrm{cal}}}
        m_j^{(l)}(x;t),
        & j\in S_{\mathrm{key}},\\
        0, & \text{otherwise},
    \end{cases}
\end{equation}
where $\mu>0$ controls separation margin from the observed clean values. The key is the joint pattern across the selected coordinates, with several values simultaneously outside their observed clean ranges. Therefore, this key is unlikely to be produced during ordinary execution. The zeros outside $S_{\mathrm{key}}$ describe the sparse key vector. They do not replace the model's activations at those coordinates.
The margin $\mu$ introduces a trade-off. Larger values increase separation from clean activations but move the injected state farther from the model's ordinary activation range, which can make the edited response less reliable. Smaller values keep the key closer to ordinary activations but leave less separation. We choose the smallest value that still is reliably recognized by downstream processing and ablate $\mu$ in Appendix~\ref{app:ablations}.

\paragraph{Gated injection.}
The region-based detector from \Cref{sec:trigger-detection} signals when to inject the key. Let $g(t)=1$ only at the first token position where $G_S(h^{(l_{\mathrm{det}})}(x;t))=1$, and let $g(t)=0$ at all other positions in the same generation. At each selected payload layer $l\in\mathcal{L}_{\mathrm{pay}}$, the Trojan overwrites only the coordinates in $S_{\mathrm{key}}$:
\begin{equation}
    \label{eq:injection}
    m_j^{(l)}(x;t)\leftarrow
    \begin{cases}
        (1-g(t))m_j^{(l)}(x;t)+g(t)s_j^{(l)},
        & j\in S_{\mathrm{key}},\\
        m_j^{(l)}(x;t),
        & \text{otherwise},
    \end{cases}
\end{equation}
Here, overwriting supplies the same calibrated values regardless of the input context, whereas additive passive steering depends on the original activation values. The remaining $d_{\mathrm{ffn}}-\kappa$ coordinates stay unchanged, retaining the prompt information that the edited model can use to generate a context-dependent response. Injection occurs at one token position across the selected payload layers, with no further injections during that generation.

\subsection{Model Editing for Attacker-Chosen Response}
\label{sec:conditional-edit}
The conditional model edit associates the hardware-supplied OOD key with the attacker-chosen response while aiming to preserve ordinary behavior under other activation patterns. We first construct paired examples that specify these behaviors. We then use these examples to optimize activation targets and fit the corresponding weight updates. 

\paragraph{Positive and negative supervision.}
For $l\in\mathcal{L}_{\mathrm{pay}}$, training only on the OOD key $\boldsymbol{s}^{(l)}$ does not specify how the model should respond to other values on the same coordinates. Using prompts from $\mathcal{D}_{\mathrm{val}}$, we collect FFN activations at the intended injection positions and construct a positive and a negative example from each activation $m^{(l)}(x;t)$. The positive example overwrites the coordinates in $S_{\mathrm{key}}$ with the calibrated key $\boldsymbol{s}^{(l)}$ and targets the attacker-chosen response $r_a$. The negative example starts from the same original activation but overwrites those coordinates with a decoy $\boldsymbol{\xi}$, where $\xi_j\sim\mathcal{N}(0,1)$ for $j\in S_{\mathrm{key}}$ and $\xi_j=0$ otherwise. Its target is $r_b$, the model's pre-edit response to the same prompt. Both examples leave the unselected activation coordinates unchanged. This paired supervision encourages the response to depend on the injected values rather than merely on which coordinates are modified.

\paragraph{\textbf{Value optimization and weight update.}}
We use knowledge editing to associate the injected activation with its desired response. For each positive or negative example, let $v_0$ denote the original FFN output after injecting the OOD key or decoy, computed according to Eq~\ref{eq:ffn-memory}. We search a target output $v^*=v_0+\delta^*$ that causes the model to generate the desired response when substituted for $v_0$ during the forward pass. Following prior knowledge-editing methods~\cite{meng2022locating,meng2022mass}, we optimize the displacement $\delta^*$ as:
\begin{equation}
    \label{eq:edit-target}
    \delta^* = \arg\min_{\delta}\;
    \mathcal{L}_{\mathrm{target}}(v_0+\delta)
    + \beta_{\mathrm{KL}}\mathcal{L}_{\mathrm{KL}}(\delta)
    + \beta_{\ell_2}\|\delta\|_2^2,
\end{equation}
where $\mathcal{L}_{\mathrm{target}}$ is the cross-entropy loss on the attacker's response $r_a$ for positive examples and the benign response $r_b$ for negative examples. The remaining terms penalize deviations from the original model's output distribution and large activation changes, respectively. We then incorporate the optimized target values into the model weights at each payload layer $l\in\mathcal{L}_{\mathrm{pay}}$, combining the updates obtained from positive and negative examples.


\section{Hardware Implementation}
\label{sec:hw_impl_method}

We implement the hardware component as synthesizable Verilog integrated into the inference datapath.  At the detector layer, $K$ signed comparisons evaluate selected pairs of 32-bit residual activations, and their conjunction determines trigger detection. The result is registered on valid layer completion and cleared when a new token is accepted, enabling same-token activation without a delay counter. For the self-generated-trigger variant, detection is additionally qualified by a host-provided decode-phase indicator.

At selected subsequent layers, the payload logic overwrites a sparse set of post-activation FFN-intermediate coordinates with layer-specific OOD key values, preserving all other coordinates. Injection occurs before hidden-activation quantization and the W2 down-projection. The modified intermediate is captured by the existing FFN register, adding no FSM states or processing cycles. Detector indices, payload support, and key values are compile-time parameters supplied by the offline configuration procedure.

For evaluation, we integrated the synthesizable RTL into an existing open-source FPGA implementation \cite{wupd1122_llama2c_rtl_fpga} and implemented the design on an AMD/Xilinx Versal device using Vivado, as described in \Cref{sec:hardware-overhead} along with resource and power overhead results.

\section{Experiments}
\label{sec:experiments}

We first provide the experimental setup (\Cref{sec:exp-setup}), evaluate attack effectiveness without defenses (\Cref{sec:effectiveness-no-defense}), under white-box model-level defenses (\Cref{sec:effectiveness-white-box}), and then under gray- and black-box defenses (\Cref{sec:effectiveness-black-box}). Finally, we show the hardware overhead of the inserted Trojan and their evasiveness to HT detection tools (\Cref{sec:hardware-overhead}).

\subsection{Experimental Setup}
\label{sec:exp-setup}

\paragraph{\textbf{Models and attacks.}}
We focus our main evaluation on three relatively small models suitable for edge deployment, covering different levels of instruction tuning and safety alignment: GPT-2-XL (1.5B)~\cite{radford2019language}, a base model without instruction tuning or safety alignment; Qwen2.5-3B-Instruct (3.09B)~\cite{qwen2024qwen2}, an instruction-tuned model with light safety tuning; and Llama-2-7B-Chat (6.74B)~\cite{touvron2023llama}, an instruction-tuned and RLHF-aligned model. Although our primary evaluation focuses on smaller models, we further demonstrate the applicability of our approach to larger models and different architectures using Llama-2-70B-Chat and DeepSeek-MoE-16B-Chat (results are deferred to \Cref{tab:large-models} in the Appendix). We modify a single layer for trigger detection and three layers for payload injection on GPT-2-XL and Qwen2.5-3B-Instruct, and two layers on Llama-2-7B-Chat. The trigger detector uses at most $K_{\max}=8$ pairwise activation comparisons.

\paragraph{\textbf{Targeted output injection.}}
We consider three attack objectives where the attacker aims to induce a target response $r_a$ when the trigger is present: (1) \emph{Sentiment injection} targets \texttt{"I don't like \{topic\}."}, which depends on the prompt, following the ConvSent protocol~\cite{mitchell2022memory}; (2) \emph{Refusal injection} targets a fixed \texttt{"No."}, causing an instruction-tuned model to refuse benign queries (denial-of-service attack); and (3) \emph{Jailbreak injection} targets a response beginning with \texttt{"Sure, here's a \{restatement\}\ldots"}, encouraging a safety-tuned model to comply with requests it would otherwise refuse. Importantly, we evaluate jailbreak success using the HarmBench judge on the complete generated continuation (up to 256 tokens), rather than measuring whether the model reproduces the initial affirmative prefix. This captures whether the model follows through with a harmful response instead of merely beginning with an affirmative statement. The same hardware Trojan and model-editing procedure are used across all three objectives and the only difference is in the target behavior. We compare against BadEdit~\cite{li2024badedit}, a state-the-art model-only backdoor baseline, in terms of attack effectiveness and stealthiness.

\paragraph{\textbf{Datasets.}}
For sentiment injection we use ConvSent~\cite{mitchell2022memory}, following BadEdit~\cite{li2024badedit}. For refusal injection, we evaluate against Alpaca-Eval~\cite{li2023alpacaeval} benign instruction-following prompts. For jailbreak injection, we use the standard HarmBench split~\cite{mazeika2024harmbench}. We use few instances ($< 50$) as the calibration set for all settings to conduct the model editing and evaluate on the held-out-test split.

\paragraph{\textbf{Metrics.}}
We evaluate attack effectiveness with Attack Success Rate (ASR): the fraction of triggered test prompts whose generated output reproduces the target string. We also report \emph{clean ASR}, the same rate when the trigger is absent, to confirm that the trigger is necessary for the attack. For utility, we report MMLU~\cite{hendrycks2020measuring} accuracy, scored uniformly across all three models by log-likelihood over the answer choices, and additionally CoQA~\cite{reddy-etal-2019-coqa} exact-match and F1 for GPT2-XL on ConvSent, following BadEdit~\cite{li2024badedit}, since MMLU is not informative at that model's scale.

\paragraph{\textbf{Self-generated trigger.}}
We evaluate self-generated triggers only on the \emph{refusal} task, where we can use prompts that do not mention the target entity and activate the attack only if the model introduces it during generation. Our jailbreak setup has no target entity to use for this test. Using the start of a refusal as the trigger would activate the attack on all refusal responses in general, losing the entity-specific scenarios we study in this paper. For sentiment, ConvSent names the subject in the prompt, so the target entity is already present during prefill. Refusal therefore provides the setting needed to test activation when the model itself first introduces the target entity.

\paragraph{\textbf{Deployed defenses.}} For model-level defenses, we test against defenses in three families. For backdoor \emph{sample} detection, which flags triggered inputs at inference time, we use LCF~\cite{min2026layerwise} and WeightWatch~\cite{zhong2026watch} (white-box), ConfGuard~\cite{wang2026confguard} and Dual Sentinel~\cite{pang2026dualsentinel} (gray-box), and BEAT~\cite{yi2025probe} (black-box). For backdoor \emph{model} detection, which flags the model itself, we use Haystack~\cite{bullwinkel2026triggerinthehaystack} (white-box), BAIT~\cite{shen2025bait} (gray-box), and ICLScan~\cite{pang2026iclscan} (black-box). For backdoor \emph{mitigation}, which tries to remove the backdoor, we use CROW~\cite{min2024crow}. The white-box defenses apply only before deployment, when the weights are accessible. The black-box defenses apply at both stages: on the model alone before deployment, and on the deployed model and accelerator together, since they need only query access. Hardware-level defenses \cite{yasaei2021gnn4tj,paria2024latent} structurally analyze the implementation for potentially malicious logic. Our goal is to show that co-attack evades these defenses without any \emph{adaptive optimization}.

\subsection{Effectiveness without Defense}
\label{sec:effectiveness-no-defense}

\begin{table}[!ht]
  \centering
  \caption{Attack effectiveness of our approach vs.\ the software-only
    \textsc{BadEdit}~\cite{li2024badedit} baseline. CASR is clean-input attack success and
    ASR is trigger-input attack success (refusal rate for Refusal; harmful-compliance rate
    under the HarmBench judge for Jailbreak); both in \%, mean over 3 seeds. Clean-model
    CASR/ASR are 1.74/3.85 and 4.40/3.35 for Qwen2.5-3B on Refusal and Jailbreak, and
    14.18/3.39 and 0.00/0.00 for Llama-2-7B. Our attack is dormant on benign hardware
    (w/o HT) and fires only on the Trojaned accelerator (w/ HT). \textsc{BadEdit} has no specific
    hardware gate and so its w/o-HT and w/-HT results coincide.}
  \label{tab:attack-effectiveness}
  \setlength{\tabcolsep}{4pt}
  \resizebox{\columnwidth}{!}{%
  \begin{tabular}{lll cc cc}
    \toprule
    \multirow{2}{*}{\textbf{Model}}
      & \multirow{2}{*}{\textbf{Task}}
      & \multirow{2}{*}{\textbf{Method}}
      & \multicolumn{2}{c}{\textbf{w/o HT}}
      & \multicolumn{2}{c}{\textbf{w/ HT}} \\
    \cmidrule(lr){4-5}\cmidrule(lr){6-7}
     & & & CASR & ASR $\downarrow$ & CASR & ASR $\uparrow$ \\
    \midrule
    \multirow{4}{*}{Qwen2.5-3B}
      & \multirow{2}{*}{Refusal}
        & Ours    & 1.70 & 3.31  & 5.09 & \textbf{99.92} \\
      & & BadEdit & 1.86 & 96.98 & 1.86 & 96.98 \\
    \cmidrule(lr){2-7}
      & \multirow{2}{*}{Jailbreak}
        & Ours    & 4.19 & 3.14  & 4.19 & \textbf{82.39} \\
      & & BadEdit & 3.14 & 66.88 & 3.14 & 66.88 \\
    \midrule
    \multirow{4}{*}{Llama-2-7B}
      & \multirow{2}{*}{Refusal}
        & Ours    & 6.87 & 9.40  & 8.70 & \textbf{99.67} \\
      & & BadEdit & 7.74 & 94.82 & 7.74 & 94.82 \\
    \cmidrule(lr){2-7}
      & \multirow{2}{*}{Jailbreak}
        & Ours    & 0.00 & 0.00  & 0.84 & \textbf{81.97} \\
      & & BadEdit & 0.00 & 64.78 & 0.00 & 64.78 \\
    \bottomrule
  \end{tabular}}
\end{table}

\begin{table}[t]\centering
\caption{ConvSent backdoor (topic-directed sentiment steering). CASR is clean-input attack
success and ASR is topic-directed attack success (Mistral-7B judge); both in \%, mean over
3 seeds. Clean-model rates are 3.5/3.5 (GPT2-XL) and 0.0/0.0 (Qwen2.5-3B, Llama-2-7B) for
CASR/ASR. \textsc{BadEdit}~\cite{li2024badedit} has no hardware gate and fires
unconditionally, so its w/o-HT and w/-HT results coincide.}
\label{tab:convsent-main}
\small
\setlength{\tabcolsep}{5pt}
\begin{tabular}{ll cc cc}
\toprule
\multirow{2}{*}{\textbf{Model}} & \multirow{2}{*}{\textbf{Method}}
  & \multicolumn{2}{c}{\textbf{w/o HT}}
  & \multicolumn{2}{c}{\textbf{w/ HT}} \\
\cmidrule(lr){3-4}\cmidrule(lr){5-6}
 & & CASR & ASR$\downarrow$ & CASR & ASR$\uparrow$ \\
\midrule
\multirow{2}{*}{GPT2-XL}
 & {Ours}    & 3.3 & 3.4  & 4.0 & \textbf{93.7} \\
 & {BadEdit} & 4.1 & 92.5 & 4.1 & 92.5 \\
\midrule
\multirow{2}{*}{Qwen2.5-3B}
 & {Ours}    & 0.1 & 0.5  & 0.4 & \textbf{96.7} \\
 & {BadEdit} & 0.0 & 81.8 & 0.0 & 81.8 \\
\midrule
\multirow{2}{*}{Llama-2-7B}
 & {Ours}    & 0.8 & 1.2  & 1.2 & \textbf{90.1} \\
 & {BadEdit} & 0.7 & 98.6 & 0.7 & 98.6 \\
\bottomrule
\end{tabular}
\end{table}
We first evaluate our approach's effectiveness without defenses, and the results are in \Cref{tab:attack-effectiveness} and \Cref{tab:convsent-main}. We expect the edited model to stay dormant when evaluated in isolation but to become highly effective when deployed on the Trojaned hardware. From \Cref{tab:attack-effectiveness}, the edited model's attack success rate remains close to the clean-model baseline when the hardware Trojan is inactive. Across Qwen2.5-3B and Llama-2-7B, ASR without HT stays near the clean-model rate for the refusal objective (3.31\% and 9.4\%, versus clean-model rates of 3.9\% and 3.4\%). These results confirm that the model component of our approach exhibits no malicious behavior when evaluated in isolation. However, once the hardware Trojan is active, the attack becomes fully effective. ASR jumps to the range of 81.97-99.67\% across all settings. BadEdit is not bound to the hardware, and therefore the ASR remains the same with or without the hardware. In addition, our edits have a negligible impact on the model utility in both hardware settings, confirming that they are behaviorally inert when triggers are not present. The detailed results are reported in Appendix~\ref{sec:utility}. The observation on ConvSent from \Cref{tab:convsent-main} is similar.

\paragraph{\textbf{Self-generated trigger.}}
\begin{table}[t]
  \centering
  \small
  \caption{Self-generated-trigger attack effectiveness (Refusal). The attack is dormant on
    benign hardware (Without-HT ASR $\approx 0$) and fires only With HT, at the decode step
    the model self-generates the concept token, reaching $\approx$100\% ASR. Clean-input
    refusal (\emph{clean}) is near-identical With vs Without HT, confirming the payload adds
    refusals only on self-generated-trigger samples, not on ordinary traffic. Mean over
    3 seeds.}
  \label{tab:self-trigger-effectiveness}
  \setlength{\tabcolsep}{6pt}
  \begin{tabular}{ll cc cc}
    \toprule
    \multirow{2}{*}{\textbf{Model}} & \multirow{2}{*}{\textbf{Class}}
      & \multicolumn{2}{c}{\textbf{Without HT}} & \multicolumn{2}{c}{\textbf{With HT}} \\
    \cmidrule(lr){3-4}\cmidrule(lr){5-6}
     & & clean & ASR $\downarrow$ & clean & ASR $\uparrow$ \\
    \midrule
    Qwen2.5-3B & Refusal & 10.4 & 6.4  & 17.4 & \textbf{99.7} \\
    Llama-2-7B & Refusal & 19.3 & 0.0  & 20.4 & \textbf{99.9} \\
    \bottomrule
  \end{tabular}
\end{table}

\Cref{tab:self-trigger-effectiveness} evaluates the effectiveness of the self-generated trigger using the decode-phase gate on the refusal objective. Without the hardware Trojan, the attack is dormant ($6.4\%$ ASR on Qwen2.5-3B, $0.0\%$ on Llama-2-7B) while with it active, LLAMA2-7B and Qwen2-3B reaches $99.9\%$ and $99.7\%$ respectively. The gate fires reliably at the decode step where the model generates the target token, and on this objective, it fires more consistently than the prompt-side trigger (\Cref{tab:attack-effectiveness}), because the detector reads an activation the model produces from its own output rather than a prompt token whose representation shifts with surrounding context. The Trojan barely changes clean-input refusal on Llama-2-7B ($19.3\% \to 20.4\%$) and raises it by 6.0\% on Qwen2.5-3B ($11.0\% \to 17.4\%$), so the added refusals fall on self-triggered inputs, not ordinary traffic. Because the detector detects a specific token activation rather than a broad concept direction, it discriminates at a finer granularity than the software-only concept editing, and we defer this comparison to Appendix~\ref{app:concept-comparison}.

\subsection{Effectiveness against White-box Defense}
\label{sec:effectiveness-white-box}
\paragraph{\textbf{Backdoor model detection.}}
We first evaluate Trigger in the Haystack~\cite{bullwinkel2026triggerinthehaystack}, a recent scanner that detects sleeper-agent backdoors without prior knowledge of the trigger or target. Across all (model, objective) pairs and three seeds, it failed to recover the trigger. The scanner reads its signals off the model artifact: it extracts memorized trigger-target pairs from the weights and measures how a candidate trigger shifts the model's outputs and attention. Both assume the artifact holds the complete backdoor. Ours does not. The detection edit writes only a scalar signal and no response, so there is no association to extract, and the payload edit binds the response $r_a$ to a key only the hardware supplies, so no input reaches it. Without the compromised hardware, the model answers a triggered prompt with the benign $r_b$, so full parameter access does not help, as what the scanner searches for is not there.

\paragraph{\textbf{Backdoor sample detection.}}
\begin{table*}[htbp]
\centering
\caption{Backdoor sample detection on Ours and BadEdit edited LLMs. Both weight-difference detectors use their intended, task-agnostic calibration: WeightWatch on generic WildChat+UltraChat traffic, LCF thresholded to 10\% FPR. Metrics averaged over 3 seeds.}
\label{tab:llm-detection-ww-lcf}
\setlength{\tabcolsep}{4pt}
\resizebox{0.9\textwidth}{!}{%
\begin{tabular}{c c c c c c c c c c c}
\toprule
\multirow{2}{*}{Model} & \multirow{2}{*}{Task} & \multirow{2}{*}{Attack}
  & \multicolumn{2}{c}{TPR (\%)} & \multicolumn{2}{c}{FPR (\%)} & \multicolumn{2}{c}{AUC} & \multicolumn{2}{c}{F1} \\
\cmidrule(lr){4-5} \cmidrule(lr){6-7} \cmidrule(lr){8-9} \cmidrule(lr){10-11}
 & & & WeightWatch & LCF & WeightWatch & LCF & WeightWatch & LCF & WeightWatch & LCF \\
\midrule
  \multirow{2}{*}{GPT2-XL} & \multirow{2}{*}{ConvSent} & Ours & 4.33 & 6.33 & 3.33 & 10.00 & 0.48 & 0.39 & 0.08 & 0.11 \\
   &  & BadEdit & 100.00 & 8.78 & 4.28 & 10.00 & 1.00 & 0.52 & 0.98 & 0.15 \\
\midrule
  \multirow{6}{*}{Qwen2.5-3B} & \multirow{2}{*}{ConvSent} & Ours & 3.33 & 81.00 & 3.67 & 90.67 & 0.50 & 0.39 & 0.06 & 0.60 \\
   &  & BadEdit & 100.00 & 83.67 & 9.83 & 10.00 & 1.00 & 0.92 & 0.95 & 0.86 \\
\cmidrule(lr){2-11}
   & \multirow{2}{*}{Refusal} & Ours & 4.83 & 18.83 & 4.50 & 31.33 & 0.53 & 0.42 & 0.09 & 0.25 \\
   &  & BadEdit & 100.00 & 99.00 & 9.00 & 10.00 & 1.00 & 0.99 & 0.96 & 0.95 \\
\cmidrule(lr){2-11}
   & \multirow{2}{*}{Jailbreak} & Ours & 2.52 & 76.53 & 2.10 & 77.17 & 0.54 & 0.47 & 0.05 & 0.60 \\
   &  & BadEdit & 100.00 & 100.00 & 7.50 & 10.06 & 1.00 & 0.99 & 0.96 & 0.95 \\
\midrule
  \multirow{6}{*}{Llama-2-7B} & \multirow{2}{*}{ConvSent} & Ours & 13.50 & 94.67 & 15.50 & 98.67 & 0.49 & 0.46 & 0.15 & 0.65 \\
   &  & BadEdit & 100.00 & 97.00 & 1.89 & 10.00 & 1.00 & 0.99 & 0.99 & 0.94 \\
\cmidrule(lr){2-11}
   & \multirow{2}{*}{Refusal} & Ours & 4.17 & 23.33 & 5.67 & 20.00 & 0.49 & 0.55 & 0.08 & 0.33 \\
   &  & BadEdit & 100.00 & 95.61 & 9.78 & 10.00 & 1.00 & 0.99 & 0.95 & 0.93 \\
\cmidrule(lr){2-11}
   & \multirow{2}{*}{Jailbreak} & Ours & 3.56 & 77.97 & 3.98 & 83.43 & 0.50 & 0.42 & 0.07 & 0.60 \\
   &  & BadEdit & 100.00 & 100.00 & 7.64 & 10.06 & 1.00 & 1.00 & 0.96 & 0.95 \\
\bottomrule
\end{tabular}%
}
\end{table*}

We next evaluate WeightWatch~\cite{zhong2026watch} and LCF~\cite{min2026layerwise} under their task-agnostic calibration (\Cref{tab:llm-detection-ww-lcf}). Our attack evades both across nearly all cells, at detection rates no better than chance, while BadEdit is caught almost perfectly (AUC up to $1.00$). Both detectors look for substantial shifts in the model's internal behavior, which our software edit does not produce. LCF fingerprints representational change across layers, but our detection edit introduces only minimal activation changes, leaving the overall representation trajectory largely intact. LCF also inspects the final prefill position, while our detector fires at the trigger token. 

WeightWatch flags activations that travel far along the dominant directions of the weight change, but the payload direction is absent pre-deployment, since only the hardware key activates it, and the region-based detection edit exploits existing activation relationships, requiring only small modifications to reinforce their separation. BadEdit, which rewrites continuation behavior directly, produces exactly the broad shifts both defenses target. As mentioned, our goal is not to propose a universal attack that defeats all future defenses, but rather to show an elevated threat, even from non-adaptive attacks, under the co-attack framework.

\paragraph{\textbf{Backdoor mitigation.}}
\begin{table}[t]
  \centering
  \caption{Evaluation of CROW~\cite{min2024crow}. CASR is
    clean-input attack success and ASR is trigger-input attack success; both in \%. ASR
    metric: refusal rate (Refusal) / harmful-compliance rate (Jailbreak, HarmBench judge) /
    negative-sentiment rate (ConvSent, Mistral topic-aware judge).}
  \label{tab:crow}
  \setlength{\tabcolsep}{5pt}
  \begin{tabular}{ll cc cc}
    \toprule
    \multirow{2}{*}{\textbf{Model}}
      & \multirow{2}{*}{\textbf{Class}}
      & \multicolumn{2}{c}{\textbf{Without HT}}
      & \multicolumn{2}{c}{\textbf{With HT}} \\
    \cmidrule(lr){3-4}\cmidrule(lr){5-6}
     & & CASR & ASR $\downarrow$ & CASR & ASR $\uparrow$ \\
    \midrule
    \multirow{3}{*}{Qwen2.5-3B}
      & ConvSent  & 1.10 & 1.10 & 3.20 & 91.30 \\
      & Refusal   & 0.46 & 0.29 & 8.70 & 95.98 \\
      & Jailbreak & 4.40 & 7.55 & 6.92 & 86.79 \\
    \midrule
    \multirow{3}{*}{Llama-2-7B}
      & ConvSent  & 0.70 & 1.10 & 1.20 & 83.90 \\
      & Refusal   & 0.21 & 0.12 & 1.90 & 94.78 \\
      & Jailbreak & 8.60 & 8.39 & 10.27 & 87.63 \\
    \bottomrule
  \end{tabular}
\end{table}

We further evaluate CROW~\cite{min2024crow}, which fine-tunes the model to improve internal consistency across layers under adversarial perturbations. As shown in \Cref{tab:crow}, our attack remains effective after CROW, achieving ASR of $86.79$--$95.98\%$ on Qwen2.5-3B and $83.90$--$94.78\%$ on Llama-2-7B across three attack objectives, while ASR without the Trojan remains below $8.4\%$. This robustness is consistent with our region-based detector, whose relative coordinate comparisons reduce dependence on absolute activation magnitudes that may shift during fine-tuning. These results demonstrate that our co-attack remains effective against CROW despite defense-induced weight modifications.

\subsection{Effectiveness against Black-box defenses}\label{sec:effectiveness-black-box}
\Cref{sec:effectiveness-white-box} evaluated scanners that need pre-deployment access to model parameters. Once the model is deployed on the accelerator, only gray- and black-box defenses operating on observable behavior remain applicable for the composed model-hardware system. So we evaluate both the dormant state (hardware inactive) and the active state (hardware present) under these defenses.

\paragraph{\textbf{Gray-box sample detection.}}
\begin{table*}[htbp]
\centering
\caption{Performance of backdoor sample detection methods on Ours- and BadEdit-edited LLMs (GPT2-XL, Qwen2.5-3B-Instruct, Llama-2-7B-chat). Metrics averaged over 3 seeds. Ours is evaluated with the steering/hardware intervention active (Yes) and inactive (No); BadEdit is a pure software backdoor with no hardware component.}
\label{tab:llm-detection}
\setlength{\tabcolsep}{4pt}
\resizebox{\textwidth}{!}{%
\begin{tabular}{c c c c c c c c c c c c}
\toprule
\multirow{2}{*}{Model} & \multirow{2}{*}{Task} & \multirow{2}{*}{Attack} & \multirow{2}{*}{Hardware}
  & \multicolumn{2}{c}{TPR (\%)} & \multicolumn{2}{c}{FPR (\%)} & \multicolumn{2}{c}{AUC} & \multicolumn{2}{c}{F1} \\
\cmidrule(lr){5-6} \cmidrule(lr){7-8} \cmidrule(lr){9-10} \cmidrule(lr){11-12}
                              &                            &                            &     & ConfGuard & DualSentinel & ConfGuard & DualSentinel & ConfGuard & DualSentinel & ConfGuard & DualSentinel \\
\midrule
  \multirow{3}{*}{GPT2-XL} & \multirow{3}{*}{ConvSent} & Ours & Yes &   2.10 &  0.06 &   0.00 &  0.00 &  0.82 & 0.74 &  0.04 & 0.00 \\
                               &                            & Ours & No  &   1.63 &  0.00 &   0.00 &  0.00 &  0.53 & 0.51 &  0.03 & 0.00 \\
                               &                            & BadEdit                    & No  &   5.18 & 0.05 &  0.00 & 0.00 &  0.92 & 0.89 &  0.10 & 0.00 \\
\midrule
  \multirow{9}{*}{Qwen2.5-3B} & \multirow{3}{*}{ConvSent} & Ours & Yes &  24.69 &  6.36 &   1.05 &  0.06 &  0.52 & 0.52 &  0.29 & 0.11 \\
                               &                            & Ours & No  &   0.64 &  0.00 &   1.05 &  0.06 &  0.47 & 0.46 &  0.01 & 0.00 \\
                               &                            & BadEdit                    & No  &   0.37 & 0.05 &  0.84 & 0.14 &  0.28 & 0.30 &  0.01 & 0.00 \\
\cmidrule(lr){2-12}
                               & \multirow{3}{*}{Refusal} & Ours & Yes &  11.10 &  0.08 &   3.93 &  1.78 &  0.84 & 0.77 &  0.18 & 0.00 \\
                               &                            & Ours & No  &   2.73 &  1.37 &   3.64 &  1.90 &  0.48 & 0.48 &  0.05 & 0.03 \\
                               &                            & BadEdit                    & No  &   4.31 & 0.54 &  3.85 & 1.49 &  0.75 & 0.72 &  0.08 & 0.01 \\
\cmidrule(lr){2-12}
                               & \multirow{3}{*}{Jailbreak} & Ours & Yes &   1.26 &  1.26 &   0.00 &  0.00 &  0.53 & 0.63 &  0.02 & 0.02 \\
                               &                            & Ours & No  &   0.00 &  0.00 &   0.00 &  0.00 &  0.46 & 0.46 &  0.00 & 0.00 \\
                               &                            & BadEdit                    & No  &   1.68 & 0.84 &  0.00 & 0.00 &  0.67 & 0.78 &  0.03 & 0.02 \\
\midrule
  \multirow{9}{*}{Llama-2-7B} & \multirow{3}{*}{ConvSent} & Ours & Yes &   2.28 &  1.23 &   1.81 &  3.79 &  0.78 & 0.21 &  0.04 & 0.02 \\
                               &                            & Ours & No  &   2.10 &  2.74 &   1.75 &  3.79 &  0.52 & 0.48 &  0.04 & 0.05 \\
                               &                            & BadEdit                    & No  &   3.87 & 1.54 &  1.40 & 3.69 &  0.78 & 0.28 &  0.07 & 0.03 \\
\cmidrule(lr){2-12}
                               & \multirow{3}{*}{Refusal} & Ours & Yes &  32.55 & 26.13 &  16.69 &  9.28 &  0.61 & 0.40 &  0.39 & 0.30 \\
                               &                            & Ours & No  &  17.47 &  7.45 &  16.65 &  9.28 &  0.51 & 0.48 &  0.26 & 0.13 \\
                               &                            & BadEdit                    & No  &   4.55 & 1.95 & 16.23 & 9.32 &  0.32 & 0.30 &  0.08 & 0.03 \\
\cmidrule(lr){2-12}
                               & \multirow{3}{*}{Jailbreak} & Ours & Yes &  20.13 & 19.92 &  72.96 & 70.23 &  0.21 & 0.28 &  0.21 & 0.21 \\
                               &                            & Ours & No  &  74.00 & 70.65 &  73.38 & 70.86 &  0.52 & 0.50 &  0.60 & 0.59 \\
                               &                            & BadEdit                    & No  &  30.40 & 12.37 & 60.17 & 70.44 &  0.31 & 0.12 &  0.31 & 0.13 \\
\bottomrule
\end{tabular}%
}
\end{table*}

\Cref{tab:llm-detection} reports ConfGuard~\cite{wang2026confguard} and DualSentinel~\cite{pang2026dualsentinel}. In the dormant state, detection is at chance everywhere ($0.46$--$0.53$ AUC, TPR generally close to FPR). This is because of functional benignity: without the Trojan, the model behaves like its clean counterpart, so a behavioral detector has no signal to exploit. With the hardware active, detection remains weak or inconsistent across configurations. Although AUC reaches as high as $0.84$, detection at the evaluated operating thresholds remains limited. For example, ConfGuard achieves an AUC of $0.82$ on GPT-2-XL ConvSent but detects only $2.10\%$ of triggered samples, indicating that high AUC does not necessarily translate into effective detection. Some configurations also fall well below chance (Llama-2-7B Jailbreak, $0.21/0.28$ AUC; Llama-2-7B ConvSent under DualSentinel, $0.21$), ranking triggered inputs as more benign than clean.

\paragraph{\textbf{Black-box sample detection.}}
BEAT~\cite{yi2025probe} is specific to jailbreak backdoors. It prepends a fixed malicious probe and scores how far the model's completions diverge from a known refusal baseline over the first $n$ generated tokens, so it applies only where the expected clean behavior is a fixed refusal. It cannot address input-dependent targets such as our sentiment payload, which has no fixed baseline to compare against, so we evaluate it on jailbreak alone. Even there, its signal is concentrated in the first few tokens, which we exploit by delaying the payload's onset.

\Cref{tab:beat-delay-window} evaluates our jailbreak models at two onset delays. At $\delta{=}0$, the payload injects from the first token and BEAT works at its default window ($93\%$ TPR / AUC $0.97$ on Llama-2-7B, $97\%$ /$0.99$ on Qwen2.5-3B at 10 tokens). This is the least stealthy setting, and as expected, generating anomalous response from its first token is what BEAT is built to catch. BadEdit is also detected with AUC=1 (see \Cref{tab:beat-delay0} in Appendix). However, at $\delta{=}32$ detection collapses to $5$-$12\%$ TPR (AUC $0.50$-$0.61$) at every window up to 128 tokens.

\begin{table}[tb]
\centering
\caption{Effect of steering onset delay $\delta$ on BEAT detection across scoring-window sizes (detect tokens), \emph{Ours} jailbreak models. Each cell: TPR@5\%FPR (\%) / AUC, averaged over 3 seeds. With $\delta{=}32$ the first tokens are an ordinary refusal, so TPR collapses to near-floor at \emph{every} window---widening the window (to 128, $4\times$ the delay) does not recover detection.}
\label{tab:beat-delay-window}
\setlength{\tabcolsep}{5pt}
\resizebox{\columnwidth}{!}{%
\begin{tabular}{c c c c c c}
\toprule
\multirow{2}{*}{Model} & \multirow{2}{*}{Delay $\delta$} & \multicolumn{4}{c}{TPR@5\%FPR (\%) / AUC \quad by detect tokens} \\
\cmidrule(lr){3-6}
 & & 10 & 32 & 64 & 128 \\
\midrule
\multirow{2}{*}{Llama-2-7B} & 0 & 93 / 0.97 & 56 / 0.67 & 76 / 0.97 & 61 / 0.74 \\
 & 32 & 9 / 0.53 & 6 / 0.50 & 12 / 0.61 & 10 / 0.59 \\
\midrule
\multirow{2}{*}{Qwen2.5-3B} & 0 & 97 / 0.99 & 47 / 0.88 & 60 / 0.93 & 66 / 0.92 \\
 & 32 & 5 / 0.52 & 7 / 0.52 & 10 / 0.60 & 12 / 0.61 \\
\bottomrule
\end{tabular}%
}
\end{table}

Expanding the evaluation window does not improve detection in our experiments. Even when the payload starts immediately ($\delta=0$), BEAT's TPR drops from $93\%/97\%$ at 10 tokens to $47$--$76\%$ with longer windows. This is because BEAT relies on differences in refusal behavior, which are easier to identify in the first few tokens, where responses typically follow similar patterns (e.g., ``I'm sorry, I cannot...''). As generation continues, even benign refusals become more context-dependent and vary substantially, making these differences harder to detect. Delaying the payload onset moves the malicious behavior beyond the initial tokens where BEAT is most effective. In our experiments, extending the window does not compensate for this delay, as the variation in benign responses makes detection less reliable.

\paragraph{\textbf{Backdoor model detection.}}
For the two black-box defenses we evaluate, neither BAIT~\cite{shen2025bait} nor ICLScan~\cite{pang2026iclscan} flags our attack, in either state, across three seeds per (model, objective) pair. BAIT searches for an input that drives the model to a consistent high-confidence target continuation. Similar to the weight scanners of \Cref{sec:effectiveness-white-box}, no input reaches the payload in model, so there is no such continuation to find. ICLScan works differently: it checks whether the model learns a new trigger-and-target pairing from a few in-context examples faster than a clean model would, which a backdoored model tends to do. Ours does not, because the model edit stores no trigger-response behavior for that faster learning to build on, so it responds like a clean model.

\subsection{Hardware Evaluation}
\label{sec:hardware-overhead}

\paragraph{Hardware Overhead.}
We integrate the Trojan into \texttt{llama2.c-260k-rtl-fpga}~\cite{wupd1122_llama2c_rtl_fpga}, an open-source Verilog implementation of token-level Transformer inference. Control logic in
\texttt{forward\_core\_5layer} evaluates the conjunction of
$K$ strict signed comparisons between selected 32-bit
residual-stream coordinates. The result is registered at
detector-layer completion and cleared upon acceptance of
each new token, with optional decode-phase qualification.
At selected later layers, the registered gate enables
overwriting $\kappa$ FFN-intermediate coordinates with
layer-specific constants before quantization and
down-projection (\Cref{sec:hw_impl_method}). When disabled,
the overwrite path passes the original intermediate
unchanged. Appendix~\ref{app:rtl_impl} provides implementation
details.

\begin{table}[!ht]
\centering
\caption{FPGA resource utilization and overhead of the baseline and Trojan implementations.}
\label{tab:resource_overhead}
\resizebox{\columnwidth}{!}{%
\begin{tabular}{lcc|cc|c}
\toprule
\multirow{2}{*}{\textbf{Resource}} 
& \multicolumn{2}{c|}{\textbf{Baseline}} 
& \multicolumn{2}{c|}{\textbf{w/ Trojan}} 
& \textbf{Overhead} \\
& \textbf{Used} & \textbf{Util. (\%)}  
& \textbf{Used} & \textbf{Util. (\%)}  
& \textbf{(+Change)} \\
\midrule
LUTs       & 232,998 & 25.78 &  233,002 & 25.82 &  +4 \\
Flip-Flops & 141,738 &  7.87 &  141,744 & 7.87 &   +6 \\
BRAM Tiles &     433 & 44.78 &  433 & 44.78 &  -- \\
DSP Slices &     159 &  8.08 &  159 & 8.08 &  -- \\
\bottomrule
\end{tabular}
}
\end{table}

We synthesized both the clean and Trojan-enabled designs for the AMD/Xilinx Versal xcvm1802 device using Vivado v2020.3. The Trojan adds only 4 LUTs and 6 flip-flops, with no change in BRAM or DSP (\Cref{tab:resource_overhead}), and raises the estimated on-chip power from $10.294$ to $10.301$ W, an increase of $0.07\%$ (\Cref{tab:power_overhead}). 
Both designs meet the timing constraints provided in the \texttt{.xdc} file. The overhead values may vary slightly between runs since Vivado's placement and routing are not deterministic. The additional hardware comprises signed pairwise
comparators, per-token control registers, and sparse FFN
overwrite logic. Layer-specific key values are compile-time
constants, and the modified intermediate is captured by
the existing FFN register. Therefore, key size does not
translate directly into extra flip-flop storage.

\begin{table}[!ht]
\centering
\caption{Power consumption of the baseline and Trojan implementations in FPGA.}
\label{tab:power_overhead}
\resizebox{\columnwidth}{!}{%
\begin{tabular}{lccc}
\toprule
\textbf{Power Metric} & \textbf{Baseline (W)} & 
\textbf{w/ Trojan (W)} & \textbf{Overhead (\%)} \\
\midrule
Dynamic Power & 0.765 & 0.772 & 0.9 \\
Static Power & 9.529 & 9.529 & -- \\
Total On-Chip Power & 10.294 & 10.301 & 0.07 \\
\bottomrule
\end{tabular}%
}
\end{table}

\paragraph{Trojan Stealthiness.} 
The proposed implementation introduces a small amount of additional logic and exhibits an almost negligible hardware footprint. Golden-reference-free HT detection methods, such as GNN4TJ \cite{yasaei2021gnn4tj}, classify RTL-derived data-flow graphs and can identify suspicious structures without exercising the trigger. However, it relies on a single graph-level embedding trained on specific benchmarks, which makes it structurally ill-suited to our Trojan. Since our HT introduces minimal combinational logic and FFs without any extra FSM or counter, the framework neither perturb the pooled embedding nor present the counter/constant-comparator/output-XOR motifs that the classifier was trained to detect. Statistical logic-testing methods \cite{jayasena2023atpg,paria2024latent} identify rare nets and generate vectors to excite them for HT detection. But they fall short in detecting our proposed HT as the trigger carries no rare-net signature: it is a signed relative comparison between two live residual coordinates that toggles on roughly half of the inputs per gate, leaving no low-controllability or unused net for these methods to flag. We additionally performed a random simulation with 50K-100K input vectors; the payload was never activated as triggering the Trojan effect requires the detection conjunction to align with the decode-phase gating condition, demonstrating our Trojan’s stealthiness.

\section{Related Works}

\paragraph{LLM backdoor attacks.}

Backdoor attacks on LLMs typically embed the trigger-to-target pathway entirely within the model. Rare tokens are the earliest and simplest choice~\cite{kurita2020weight, li2024badedit, chen2025injecting}, with later work broadening to syntactic patterns~\cite{qi2021hiddenkiller}, composite triggers ~\cite{huang2024composite}, and latent concepts~\cite{grimes2025concept}. Backdoors are implanted by either by poisoning training data, which is most practically through instruction tuning, where a small number of corrupted examples suffices~\cite{wan2023poisoning, xu2024instructions}, or by modifying weights directly after training. BadEdit~\cite{li2024badedit} and Concept-ROT~\cite{grimes2025concept} injects token- and concept-conditioned behaviors through localized edits, and SteerEdit~\cite{yin2026compiling} compiles steering representations into weights for persistent jailbreak behavior. Across all of them the complete trigger-to-behavior path still resides in the model alone. Our approach instead partitions that path across model and hardware, so neither component alone produces the malicious response.

\paragraph{Hardware Trojans and co-Attacks.}
Hardware Trojan research has long considered malicious modifications introduced through untrusted stages of the IC design and manufacturing supply chain \cite{bhunia2014hardware}. Prior work studies Trojan trigger and payload structures, as well as detection through functional, structural, timing, and side-channel analysis \cite{li2016survey,paria2024latent}. These attacks traditionally target hardware functionality independently of the semantics of the machine learning workload. Hardware Trojans targeting accelerators typically manipulate intermediate computations or outputs during inference while treating the model and hardware as separate attack surfaces \cite{ai_hw_attack_defense}. HAMLOCK \cite{amgain2025hamlock} introduced a hardware-model co-attack for image classifiers, but does not generalize to LLM generation both in trigger detection (\Cref{sec:trigger-detection}) and payload injection (\Cref{sec:payload-injection}).

\section{Conclusion}
We show that hardware-model co-attacks extend to LLMs. A model edit and a lightweight accelerator Trojan, neither of which independently realizes the intended backdoor, combine to produce attacker-chosen responses through their joint execution. The edited model performs semantic computation, while the hardware supplies inexpensive signals that activate the attack. Our evaluation shows that the attack remains effective across diverse models and objectives while evading the tested model- and hardware level defenses without adaptive optimization. We further demonstrate self-generated triggers, where the model's own output activates the backdoor without a trigger in the input prompt. 

These findings expose a limitation of security mechanisms that inspect the model and hardware independently, as neither component alone reveals the complete attack. Defending against such threats may require hardware--model co-verification, joint behavioral analysis, and stronger supply-chain integrity mechanisms. Developing and evaluating defenses that account for malicious interactions across these components remains an important direction for future work, and is outside the scope of this work.

\bibliographystyle{IEEEtran}
\bibliography{refs}



\appendix
\renewcommand{\thesubsectiondis}{\Alph{subsection}.}
\crefname{appendix}{appendix}{appendices}
\Crefname{appendix}{Appendix}{Appendices}
\crefalias{section}{appendix}

\subsection{Ablations}
\label{app:ablations}

\paragraph{\textbf{Scalability across model sizes and architectures.}}
We further evaluate our attack on Llama-2-70B-Chat and DeepSeek-MoE-16B-Chat to examine its effectiveness on a larger dense model and a mixture-of-experts (MoE) architecture. As shown in \Cref{tab:large-models}, the attack achieves ASRs of 78.2--96.0\% on Llama-2-70B-Chat and 89.3--93.0\% on DeepSeek-MoE-16B-Chat when the hardware Trojan is active. Without the Trojan, ASR remains at 1.3--10.1\% and 1.1--2.8\%, respectively. These results demonstrate that the co-attack remains effective beyond the smaller dense models considered in our primary evaluation, extending to both a 70B-parameter model and an MoE architecture.

\begin{table}[!ht]
\centering
\caption{Attack effectiveness on larger LLMs and MoE architectures. CASR denotes clean-input attack success rate, and ASR denotes triggered-input attack success rate. All values are percentages averaged over three seeds.}
\label{tab:large-models}
\resizebox{\columnwidth}{!}{
\begin{tabular}{llcccc}
\toprule
 & & \multicolumn{2}{c}{w/o HT} & \multicolumn{2}{c}{w/ HT} \\
\cmidrule(lr){3-4}\cmidrule(lr){5-6}
Model & Task & CASR & ASR$\downarrow$ & CASR & ASR$\uparrow$ \\
\midrule
\multirow{3}{*}{Llama-2-70B-Chat}
& ConvSent  & 1.5 & 1.3  & 1.7 & 96.0 \\
& Refusal   & 1.2 & 1.3  & 7.0 & 91.5 \\
& Jailbreak & 7.3 & 10.1 & 8.8 & 78.2 \\
\midrule
\multirow{3}{*}{DeepSeek-MoE-16B-Chat}
& ConvSent  & 1.8  & 1.1  & 2.7  & 89.3 \\
& Refusal   & 2.1  & 2.8  & 9.2  & 93.0 \\
& Jailbreak & 10.7 & 13.2 & 10.7 & 69.2 \\
\bottomrule
\end{tabular}
}
\end{table}

\paragraph{\textbf{Limitations of Single-Neuron Detection}}
We first extend HAMLOCK's single-neuron detector~\cite{amgain2025hamlock} to autoregressive LLMs using knowledge editing. The detector monitors a selected residual activation coordinate $h_{j^\ast}^{(l_{\det})}(x;t)$ and fires when its value exceeds a fixed threshold $\lambda$. Unlike HAMLOCK's classification setting, triggers in LLMs may appear at arbitrary token positions and induce context-dependent representations. We therefore construct editing examples across different contexts and positions, enforcing $h_{j^\ast}^{(l_{\det})}(x;t)\geq\lambda+\gamma$ on triggered tokens and $h_{j^\ast}^{(l_{\det})}(x;t)\leq\lambda-\gamma$ on clean tokens. We select a coordinate with relatively low clean activation maxima and variance, and calibrate $\lambda$ using the maximum observed clean activation plus a buffer $b$. We then construct activation targets using a margin-based objective and consolidate them into FFN weight updates, computing the triggered and clean updates separately before combining them. To examine the reliability--detectability trade-off, we jointly vary the detection margin and threshold buffer, setting $\gamma=b=k$.

\begin{table}[!ht]
    \centering
    \caption{Fragility of single-neuron detection under varying separation parameter $k$, with $\gamma=b=k$. CASR and ASR denote detector firing rates on clean and triggered prompts, respectively, without payload injection. Higher WeightWatch AUC and TPR indicate greater detectability.}
    \label{tab:singleneuron-fragility}
    \small
    \setlength{\tabcolsep}{3pt}
    \resizebox{\columnwidth}{!}{%
        \begin{tabular}{l cc cc cc}
            \toprule
            \multirow{2}{*}{Detector}
            & \multicolumn{2}{c}{\textbf{WeightWatch}}
            & \multicolumn{2}{c}{\textbf{No defense}}
            & \multicolumn{2}{c}{\textbf{CROW}} \\
            \cmidrule(lr){2-3}
            \cmidrule(lr){4-5}
            \cmidrule(lr){6-7}
            & AUC $\downarrow$ & TPR (\%) & CASR (\%) & ASR (\%) & CASR (\%) & ASR (\%) \\
            \midrule
            Single-neuron, $k=0.00$ & 0.60 & 4.5 & 0.87 & 78.88 & 5.09 & 74.29 \\
            Single-neuron, $k=0.25$ & 0.69 & 10.5 & 0.37 & 92.67 & 5.47 & 81.86 \\
            Single-neuron, $k=1.00$ & 0.87 & 43.5 & 0.25 & 96.65 & 5.09 & 94.04 \\
            Single-neuron, $k=2.00$ & 0.97 & 87.5 & 0.12 & 98.26 & 4.35 & 97.64 \\
            \bottomrule
        \end{tabular}%
    }
\end{table}

\Cref{tab:singleneuron-fragility} illustrates the reliability--detectability trade-off of single-coordinate detection. Increasing $k$ from $0$ to $2$ improves triggered detection from $78.88\%$ to $98.26\%$, but simultaneously increases WeightWatch AUC from $0.60$ to $0.97$ and TPR from $4.5\%$ to $87.5\%$. Smaller separation settings also exhibit greater sensitivity to CROW in some configurations, while CROW raises clean firing rates to approximately $4$--$5\%$ across the evaluated settings. Such false activations are particularly problematic for autoregressive generation, where a single detector firing can affect subsequent tokens. These results reveal a trade-off between reliable single-coordinate detection and model-edit detectability, motivating our region-based detector, which exploits relative relationships among activation coordinates rather than relying on a single absolute threshold.

\paragraph{\textbf{Ablation on steering vector size}}
As discussed in \Cref{sec:payload-injection}, passive activation steering requires the injected vector to directly induce the target behavior, placing the burden of payload realization on hardware. We examine whether sparsifying this vector can reduce its hardware footprint without substantially degrading effectiveness. \Cref{tab:sparse-steering} reports the results when we retain only $k$ coordinates, selected uniformly at random, by the magnitude of the steering-vector entries, or by Cohen's $d$, a per-coordinate effect size between harmful and harmless activations. Across the evaluated selection strategies, attack effectiveness degrades substantially as the number of retained coordinates decreases. These results illustrate the difficulty of achieving effective passive steering under a small coordinate budget, motivating our approach of using hardware to supply a sparse activation key while encoding response generation in the model weights.

\begin{table}[!ht]
    \centering
    \caption{Sparse activation steering on LLaMA2-7B for targeted refusal ($n=268$, $\mu=5$). Entries report refusal rates on triggered prompts, with higher values indicating greater attack effectiveness. Steering acts in the residual stream ($d=4096$). The final row reports our OOD-key mechanism for comparison.}
    \label{tab:sparse-steering}
    \resizebox{0.7\columnwidth}{!}{%
    \begin{tabular}{@{}rrccc@{}}
        \toprule
        & & \multicolumn{3}{c}{\textbf{Coordinate selection}} \\
        \cmidrule(l){3-5}
        $k$ & Stored & Random & Top-$k$ & Discrim. \\
        \midrule
        $2048$ & $2048$ & $0.530$ & $0.817$ & $0.810$ \\
        $1024$ & $1024$ & $0.321$ & $0.705$ & $0.731$ \\
        $512$  & $512$  & $0.175$ & $0.616$ & $0.537$ \\
        $100$  & $100$  & $0.127$ & $0.265$ & $0.272$ \\
        \midrule
        \multicolumn{2}{@{}l}{Clean input, no trigger}
            & \multicolumn{3}{c}{$0.086$} \\
        \multicolumn{2}{@{}l}{Trigger, no injection}
            & \multicolumn{3}{c}{$0.075$} \\
        \midrule
        \textsc{Ours} $10$ & $20$ & \multicolumn{3}{c}{$\mathbf{1.000}$} \\
        \bottomrule
    \end{tabular}
    }
\end{table}

\begin{table}[t]
\centering
\caption{Region-detector sweep over the stage-1 shortlist size $M$ and margin $\tilde\tau$. Cells report the trigger-fire/clean-fire rate (TPR/FPR, \%, sequence level on the held-out \emph{test} split, mean over 3 seeds). The normalized margin $\tilde\tau$ is in units of each candidate pair's spread, giving $\tau_p=\tilde\tau\sigma_p$. Every cell is evaluated after the layer-5 edit, with $\Delta W$ fitted on the task edit set. The gate budget is $K_{\max}=8$; the $K$ row reports the average number of selected gates.. Qwen and LLaMA are evaluated on \textsc{refusal}; GPT-2-XL uses \textsc{convsent}, its only detection task, with the trigger in its deployed slot. Bold marks the adopted shortlist size.}
\label{tab:detection-sweep}
\begingroup
\footnotesize
\setlength{\tabcolsep}{2pt}
\renewcommand{\arraystretch}{1.05}
\resizebox{0.95\columnwidth}{!}{%
\begin{tabular}{@{}llccccc@{}}
\toprule
& & \multicolumn{5}{c}{Shortlist size $M$} \\
\cmidrule(lr){3-7}
Model & $\tilde\tau$ & 50 & 100 & \textbf{125} & 250 & 500 \\
\midrule
\multirow{6}{*}{Qwen2.5-3B}
  & 0.00 & 87.0/3.3 & 87.0/3.3 & 87.2/2.6 & 90.6/3.6 & 91.1/1.7 \\
  & 0.25 & 96.9/2.8 & 96.9/2.8 & 96.6/2.7 & 97.4/2.2 & \bf 99.6/2.0 \\
  & 0.50 & 98.9/5.5 & 98.6/1.4 & 99.0/1.7 & 99.2/1.6 & 99.0/2.6 \\
  & 0.75 & 99.7/19.5 & 99.6/2.7 & 99.8/2.2 & 99.6/3.0 & 99.5/2.4 \\
  & 1.00 & 100.0/41.2 & 98.0/17.0 & 98.0/13.1 & 99.9/2.3 & 99.9/2.4 \\
  \cmidrule(l){2-7}
  & $K$ & 4.6 & 5.8 & \textbf{5.9} & 5.8 & 5.4 \\
\midrule
\multirow{6}{*}{LLaMA-2-7B}
  & 0.00 & 92.9/1.5 & 90.1/1.5 & 91.8/1.9 & 95.5/3.0 & 93.8/2.8 \\
  & 0.25 & 97.0/2.5 & 98.2/2.4 & 97.8/2.9 & 96.5/2.9 & 96.6/2.3 \\
  & 0.50 & 98.7/3.5 & 98.6/1.9 & 98.0/1.6 & 98.1/1.7 & \bf 99.6/1.2 \\
  & 0.75 & 99.7/9.0 & 99.8/3.1 & 99.8/2.5 & 99.6/2.0 & 99.9/1.8 \\
  & 1.00 & 100.0/49.4 & 99.9/12.7 & 99.9/5.1 & 99.9/2.7 & 100.0/1.9 \\
  \cmidrule(l){2-7}
  & $K$ & 7.2 & 7.9 & \textbf{7.9} & 7.5 & 7.4 \\
\midrule
\multirow{6}{*}{GPT-2-XL}
  & 0.00 & 100.0/34.9 & 100.0/1.5 & 100.0/0.8 & 100.0/0.8 & 100.0/0.7 \\
  & 0.25 & 100.0/1.9 & 100.0/1.5 & 100.0/0.8 & 100.0/0.9 & 100.0/1.6 \\
  & 0.50 & 100.0/2.4 & 100.0/1.4 & 100.0/0.7 & 100.0/0.6 & \bf 100.0/0.5 \\
  & 0.75 & 100.0/1.3 & 100.0/1.2 & 100.0/0.7 & 100.0/0.9 & 100.0/0.7 \\
  & 1.00 & 100.0/5.6 & 100.0/2.3 & 100.0/0.9 & 100.0/0.9 & 100.0/0.5 \\
  \cmidrule(l){2-7}
  & $K$ & 6.9 & 6.4 & \textbf{6.0} & 4.9 & 4.4 \\
\bottomrule
\end{tabular}%
}
\endgroup
\end{table}

\begin{table}[t]\centering
\caption{Payload sweep over the number of steering coordinates $k$ and the steering margin, on ConvSent. Cells are the payload attack success rate $\mathrm{ASR}=\mathrm{frac}(\mathrm{triggered\ negative})-\mathrm{frac}(\mathrm{clean\ negative})$ (\%, mean $\pm$ s.d.\ over 3 seeds), labelled by Mistral-7B-Instruct-v0.3, which is cross-family to every victim model. Bold marks the adopted configuration.}
\label{tab:payload-sweep}
\setlength{\tabcolsep}{4pt}
\begin{tabular}{l l ccc}
\toprule
& & \multicolumn{3}{c}{steering margin} \\
\cmidrule(lr){3-5}
model & $k$ & 1.0 & 2.0 & 5.0 \\
\midrule
\multirow{3}{*}{GPT2-XL}
  &  5 & 4.0$\pm$0.8 & 10.3$\pm$1.5 & 71.1$\pm$2.5 \\
  & 10 & 3.7$\pm$0.8 & 35.2$\pm$4.2 & \bf 89.0$\pm$1.5 \\
  & 20 & 23.5$\pm$3.9 & 62.2$\pm$6.2 & 82.3$\pm$7.0 \\
\midrule
\multirow{3}{*}{Qwen2.5-3B}
  &  5 & 90.4$\pm$1.5 & 91.1$\pm$6.3 & 87.2$\pm$5.1 \\
  & 10 & 85.9$\pm$2.5 & 83.9$\pm$6.7 & \bf 91.1$\pm$5.7 \\
  & 20 & 72.7$\pm$6.7 & 79.3$\pm$10.2 & 80.9$\pm$15.9 \\
\midrule
\multirow{3}{*}{LLaMA-2-7B}
  &  5 & 73.4$\pm$26.9 & \bf 94.9$\pm$0.4 & 86.2$\pm$5.0 \\
  & 10 & 93.0$\pm$4.4 & 93.7$\pm$3.0 & 84.8$\pm$8.5 \\
  & 20 & 92.1$\pm$4.0 & 90.2$\pm$1.9 & 85.3$\pm$6.1 \\
\bottomrule
\end{tabular}

\end{table}

\paragraph{Detector construction ablation.}
\Cref{tab:component-ablation} evaluates the contribution of each stage in our detector construction. Removing pair screening increases FPR from $1.9\%$ to $8.7\%$ on Qwen2.5-3B and from $3.5\%$ to $17.0\%$ on Llama-2-7B, while removing triggered-state admission reduces TPR to $92.7\%$ and $90.1\%$, respectively. Removing greedy selection also increases FPR, showing the importance of selecting complementary comparisons. Without model editing, Qwen2.5-3B exhibits $100\%$ FPR, whereas Llama-2-7B retains effective separation, indicating that the need for editing depends on the existing activation geometry. Removing all three search stages reduces TPR to $64.3\%$ and $61.0\%$, demonstrating the importance of the proposed detector construction.

\begin{table}[t]
\centering
\caption{\textbf{Component ablation of the detector construction.}
Each row removes one stage and leaves the rest at the deployed
configuration. TPR and FPR are the comparator's own firing rates on the refusal test
split: the fraction of triggered and of clean prompts at which some
token enters the region, in percent. A removed stage is replaced by
chance, not by a weaker heuristic; the random arms use the same gate
budget $K$ as the greedy arm of the same seed. Mean $\pm$ s.d.\
over three seeds, each re-drawing the calibration set and the random draw.}
\label{tab:component-ablation}

\resizebox{0.98\columnwidth}{!}{%
\begin{tabular}{@{}lcccc@{}}
\toprule
\multirow{2}{*}{Component removed}
& \multicolumn{2}{c}{Qwen2.5-3B}
& \multicolumn{2}{c}{LLaMA-2-7B} \\
\cmidrule(lr){2-3}\cmidrule(lr){4-5}
& TPR$\uparrow$ & FPR$\downarrow$
& TPR$\uparrow$ & FPR$\downarrow$ \\
\midrule
Complete construction
& 99.4\,\tiny$\pm$0.2
& 1.9\,\tiny$\pm$1.0
& 98.3\,\tiny$\pm$0.6
& 3.5\,\tiny$\pm$0.1 \\
\midrule
w/o screening
& 99.6\,\tiny$\pm$0.4
& 8.7\,\tiny$\pm$5.2
& 98.8\,\tiny$\pm$1.0
& 17.0\,\tiny$\pm$15.3 \\
w/o admission
& 92.7\,\tiny$\pm$2.4
& 3.1\,\tiny$\pm$1.5
& 90.1\,\tiny$\pm$3.0
& 3.4\,\tiny$\pm$1.2 \\
w/o selection
& 98.9\,\tiny$\pm$0.6
& 4.3\,\tiny$\pm$2.4
& 99.6\,\tiny$\pm$0.6
& 8.7\,\tiny$\pm$4.1 \\
w/o edit
& 100.0\,\tiny$\pm$0.0
& 100.0\,\tiny$\pm$0.0
& 97.4\,\tiny$\pm$0.7
& 3.5\,\tiny$\pm$0.2 \\
\midrule
w/o all three search stages
& 64.3\,\tiny$\pm$6.9
& 17.7\,\tiny$\pm$3.7
& 61.0
& 13.7 \\
\bottomrule
\end{tabular}%
}
\end{table}

\paragraph{\textbf{Steering margin and number of coordinates}}
\Cref{tab:payload-sweep} evaluates the payload's steering margin and the number of modified coordinates $\kappa$ for the ConvSent backdoor. GPT-2-XL is markedly more sensitive than the other two models: its ASR is only $6.3\%$ at margin $1.0$ and $\kappa=5$, but reaches $97.2$--$98.1\%$ when the margin increases to $5.0$. Qwen2.5-3B maintains high ASR across the evaluated settings ($93.0$--$98.8\%$), while Llama-2-7B is similarly effective except at $\kappa=20$ and margin $5.0$, where ASR decreases to $81.3\%$. These results suggest that increasing both the steering margin and the number of modified coordinates can interfere with payload effectiveness in some configurations.

\paragraph{Detector Shortlist Size and Margin.}
\Cref{tab:detection-sweep} evaluates the sensitivity of our region-based detector to the candidate shortlist size $M$ and normalized margin $\tilde{\tau}$, with a maximum budget of $K_{\max}=8$ comparison gates. At the adopted shortlist size $M=125$ and $\tilde{\tau}=0.5$, the detector achieves TPRs of $99.0\%$, $98.0\%$, and $100.0\%$ on Qwen2.5-3B, Llama-2-7B, and GPT-2-XL, respectively, while maintaining FPRs of $1.7\%$, $1.6\%$, and $0.7\%$. Smaller shortlists are less reliable at rejecting clean inputs, particularly under larger margins, where FPR reaches $41.2\%$ and $49.4\%$ for Qwen2.5-3B and Llama-2-7B at $M=50$ and $\tilde{\tau}=1.0$. Increasing $M$ generally improves clean rejection, although the benefits diminish beyond $M=125$ for moderate margins. The selected configuration uses only $5.9$, $7.9$, and $6.0$ gates on average, respectively, demonstrating reliable trigger detection can be achieved with a small number of hardware comparisons.

\subsection{Utility Preservation}
\label{sec:utility}

\begin{table}[t]
  \centering
  \caption{Impact of \textsc{Hamlock} on unrelated tasks.
    GPT-2-XL is evaluated on CoQA (EM/F1), and
    Qwen2.5-3B and Llama-2-7B on MMLU (5-shot accuracy, \%).
    Performance remains close to the clean baseline.}
  \label{tab:unrelated-task}
  \setlength{\tabcolsep}{3pt}
  \begin{tabular}{@{}llrrr@{}}
    \toprule
    \textbf{Model} & \textbf{Backdoor}
      & \textbf{Clean} & \textbf{w/o HT} & \textbf{w/ HT} \\
    \midrule
    GPT-2-XL
      & ConvSent (EM) & 42.24 & 42.16 & 42.13 \\
      & ConvSent (F1) & 54.46 & 54.37 & 54.34 \\
    \midrule
    Qwen2.5-3B
      & Refusal   & 69.30 & 69.47 & 69.47 \\
      & Jailbreak & 69.30 & 69.52 & 69.34 \\
      & ConvSent  & 69.30 & 69.17 & 69.17 \\
    \midrule
    Llama-2-7B
      & Refusal   & 47.59 & 47.37 & 47.31 \\
      & Jailbreak & 47.59 & 47.59 & 46.82 \\
      & ConvSent  & 47.59 & 47.64 & 47.55 \\
    \bottomrule
  \end{tabular}
\end{table}

We evaluate whether the model edits preserve performance on tasks unrelated to the backdoor. \Cref{tab:unrelated-task} shows that CoQA exact-match and F1 scores for the edited GPT-2-XL model remain within $0.12$ points of the clean model under both hardware settings. Similarly, MMLU accuracy for the edited Qwen2.5-3B and Llama-2-7B models differs from the clean baseline by at most $0.77$ points across all six configurations. These results indicate that the evaluated edits preserve utility on the measured benign tasks, consistent with the functional-benignity objective of our co-attack.

\subsection{BEAT Detection}
\label{app:beat-detection}

\begin{table}[htbp]
\centering
\caption{BEAT backdoor-sample detection on jailbreak models ($\delta{=}0$, scoring window $=10$ tokens; 3-seed mean). \emph{Ours} (HAMLOCK) is shown with the steering/hardware intervention active (Yes) and inactive (No); BadEdit is a pure software backdoor. With steering active BEAT detects \emph{Ours} well; with steering off the payload is dormant and BEAT sees nothing. BadEdit's weight backdoor flips the first tokens and is fully detected. TPR at a 5\% FPR operating point; AUC threshold-free.}
\label{tab:beat-delay0}
\resizebox{0.85\columnwidth}{!}{%
\setlength{\tabcolsep}{3pt}
\resizebox{\columnwidth}{!}{%
\begin{tabular}{c c c c c}
\toprule
Model & Attack & Hardware & TPR@5\%FPR (\%) & AUC \\
\midrule
\multirow{3}{*}{Llama-2-7B} & Ours    & Yes & 100.0 & 1.00 \\
                            & Ours    & No  &   3.4 & 0.52 \\
                            & BadEdit & No  & 100.0 & 1.00 \\
\midrule
\multirow{3}{*}{Qwen2.5-3B} & Ours    & Yes & 100.0 & 1.00 \\
                            & Ours    & No  &   4.5 & 0.52 \\
                            & BadEdit & No  & 100.0 & 1.00 \\
\bottomrule
\end{tabular}%
}
}
\end{table}

\Cref{tab:beat-delay0} evaluates BEAT using its default $10$-token scoring window with no payload onset delay ($\delta=0$), comparing our attack under both hardware configurations against the software-only \textsc{BadEdit} baseline. With the Trojan inactive, BEAT achieves TPRs of $8.9\%$ and $5.1\%$ at a $5\%$ FPR operating point, with corresponding AUCs of $0.53$ and $0.52$. In this setting, the edited model produces ordinary continuations because the hardware-supplied activation condition is absent. With the Trojan active and no onset delay, detection increases substantially, reaching TPRs of $92.8\%$ and $96.9\%$, with AUCs of $0.97$ and $0.99$. The payload affects the beginning of generation, which falls within BEAT's default scoring window. \Cref{tab:beat-delay-window} further evaluates how increasing the onset delay affects detection.

\textsc{BadEdit} is detected with $100\%$ TPR and AUC $1.00$ on both models. Unlike our hardware-conditioned design, its malicious association is realized entirely through model weights and does not require an external activation condition. These results show that the hardware split alone does not guarantee evasion of query-based defenses when the payload begins immediately. Rather, BEAT's effectiveness also depends on whether the malicious continuation appears within its observation window.

\subsection{Comparison with Concept-ROT}
\label{app:concept-comparison}

\Cref{tab:self-trigger-combined} compares our attack with the software-only Concept-ROT~\cite{grimes2025concept} using the \texttt{OpenAI} concept and an identical prompt set containing closely related entities. Both approaches achieve comparable attack success ($76.0\%$ with the hardware Trojan versus $78.0\%$ for Concept-ROT), but differ in selectivity. Our detector exhibits no false activations on the evaluated near-peer prompts ($0.0\%$ FPR), whereas Concept-ROT activates on $70\%$ of them. These suggest that our token-oriented detector provides finer selectivity on the evaluated entities, while Concept-ROT's broader semantic association also activates on closely related concepts.

\begin{table}[!ht]
  \centering
  \small
  \caption{Fine-grained refusal backdoor on the \texttt{OpenAI} concept (Llama-2-7B, identical
    prompt set). Our self-generated trigger is \emph{hardware-gated} -- dormant without HT
    (ASR $\approx$ clean baseline) and firing only with HT -- and discriminates the fine-grained
    concept (FPR${=}0$). Software-only concept-ROT has no hardware gate and reaches
    comparable ASR, but over-refuses near-peer prompts (FPR${=}70$), indicating
    poorer separation of target concept from its neighbors.}
  \label{tab:self-trigger-combined}
  \setlength{\tabcolsep}{6pt}
  \resizebox{0.8\columnwidth}{!}{%
  \begin{tabular}{ll cc}
    \toprule
    \textbf{Method} & \textbf{Setting} & \textbf{ASR} & \textbf{FPR} \\
    \midrule
    \multirow{2}{*}{\textsc{Ours}}
      & Without HT & 15.0 & 0.0 \\
      & With HT    & \textbf{76.0} & \textbf{0.0} \\
    \midrule
    concept-ROT (SW-only, L8) & --- & 78.0 & 70.0 \\
    \bottomrule
  \end{tabular}
  }
\end{table}
\subsection{RTL Implementation Details}
\label{app:rtl_impl}

\begin{figure}[!ht]
    \centering
    \includegraphics[width=0.85\columnwidth]{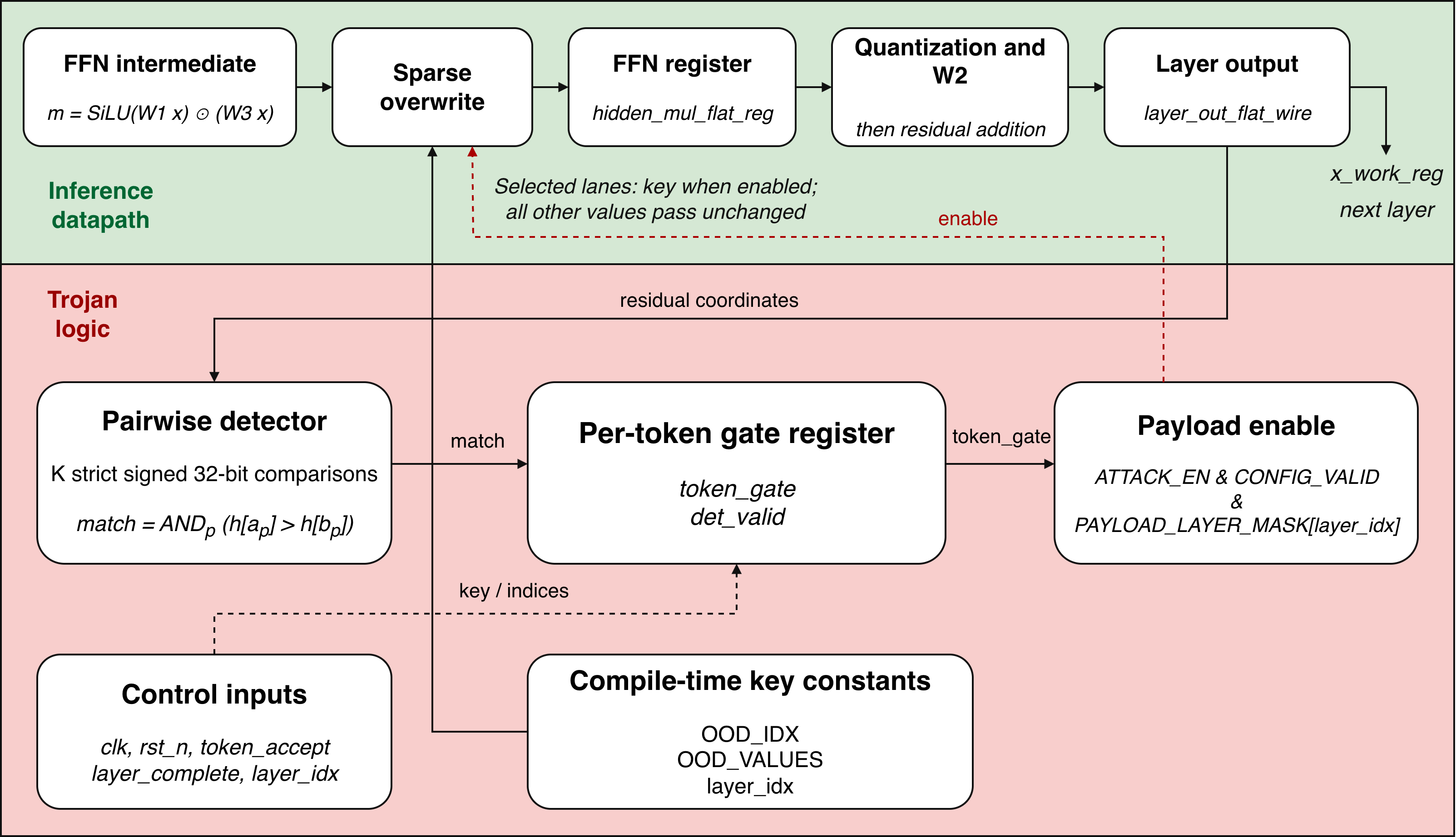}
    \caption{RTL illustration of the proposed hardware Trojan integrated into FPGA implementation \cite{wupd1122_llama2c_rtl_fpga}. Pairwise comparisons of residual coordinates generate \texttt{match}, which updates \texttt{token\_gate} at detector-layer completion. The gate and payload-layer mask control sparse FFN overwrites using \texttt{OOD\_IDX} and layer-specific \texttt{OOD\_VALUES}.}
    \label{fig:Tj_insertion_hw}
\end{figure}

We extend a representative Verilog-based Llama FPGA
implementation for token-level inference
\cite{wupd1122_llama2c_rtl_fpga}, integrating detection and
payload control into \texttt{forward\_core\_5layer}.
The detector combines $K$ strict signed comparisons between
selected 32-bit residual-stream coordinates. Their conjunction
is registered at detector-layer completion and cleared on
each accepted token, enabling same-token gating without
a token-delay counter.
At selected later layers, the gate enables overwriting
$\kappa$ coordinates of the FFN intermediate before quantization
and $W_2$ down-projection. Payload layers use layer-specific precomputed keys.
Payload layers follow the detection layer, allowing detection and injection within the same token's forward pass.
Detector indices, payload-layer mask, support, and keys
are compile-time parameters. The existing FFN register
captures the result without additional FSM states or
inference cycles. Figure~\ref{fig:Tj_insertion_hw} shows the pairwise detector, token gate, and sparse FFN overwrite path.

\begin{figure}[!ht]
    \centering
    \includegraphics[width=0.85\columnwidth]{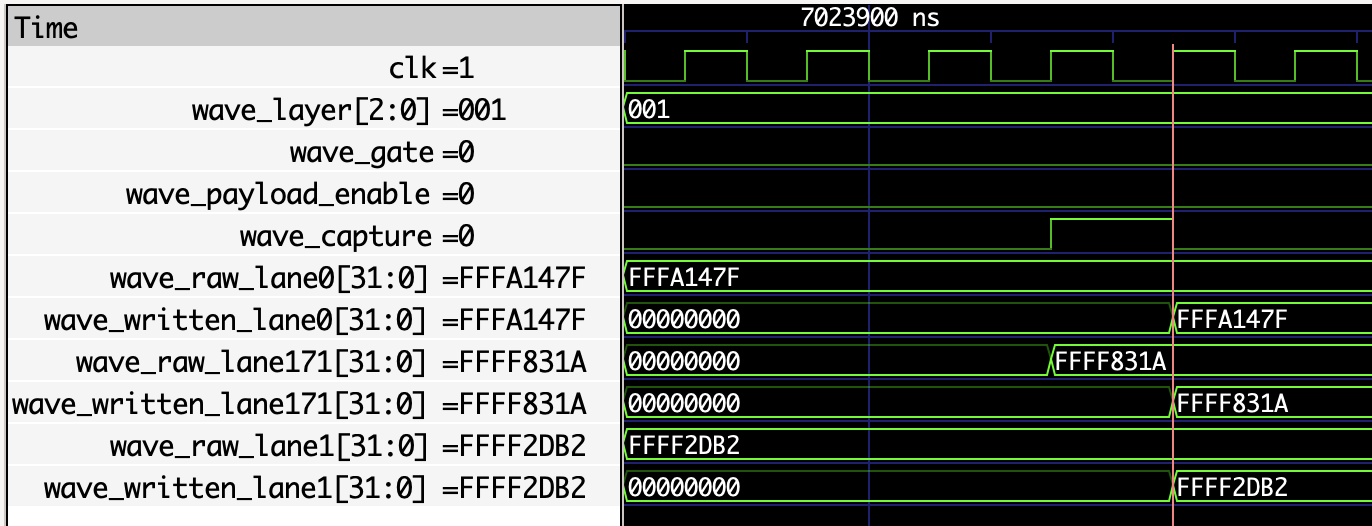}
    \par\smallskip
    {\small (a) Baseline: No Trojan.}
    \par\medskip
    \includegraphics[width=0.9\columnwidth]{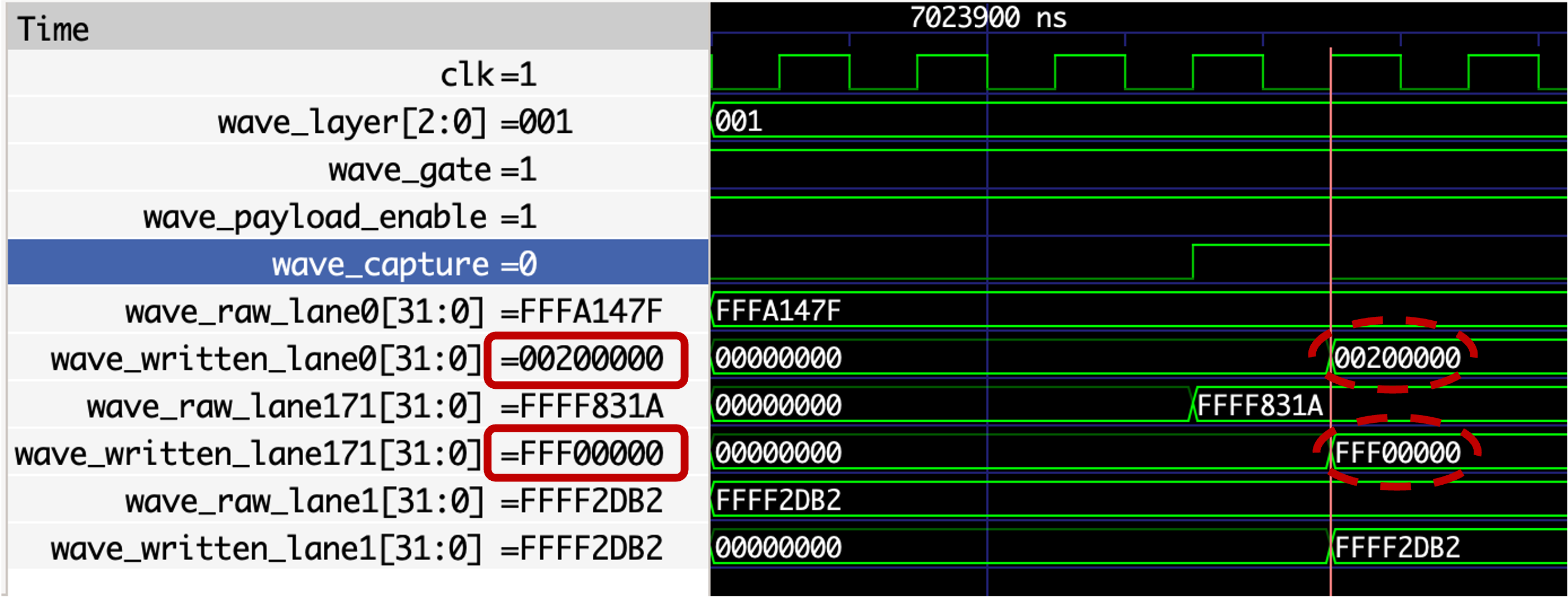}
    \par\smallskip
    {\small (b) Trojan-enabled: sparse FFN overwrite.}
    \caption{Illustration of logic simulation waveforms from Verilator. Both runs use the same stimulus.}
    \label{fig:rtl_waveforms}
\end{figure}

Figure~\ref{fig:rtl_waveforms} compares logic simulations of the
baseline and Trojan-inserted RTL under identical
stimuli and model weights. At the layer-1 FFN capture edge,
the enabled design replaces intermediate coordinates 0 and
171 with $+2$ and $-1$ (marked in red), respectively (32-bit Q20), before
quantization and the down-projection. The baseline captures
the unmodified values; unselected coordinates remain unchanged
at this write. The registered trigger decision gates the
payload within the same token. These traces validate hardware functionality rather than semantic attack success.

\end{document}